\documentclass{nature}

\usepackage{graphicx}
\usepackage{amssymb}
\usepackage{hyperref}
\usepackage{xcolor}
\usepackage{lineno}

\title{\large Young asteroid families as the primary source of meteorites}

\author{M.~Bro\v z$^1$, P.~Vernazza$^2$, M. Marsset$^{3,4}$, F.E. DeMeo$^4$, R.P. Binzel$^4$, D. Vokrouhlick\'y$^1$, D.~Nesvorn\'y$^5$}

\newcommand{\aap}{Astron.\ Astrophys.}

\newcommand{\apjl}{Astrophys.\ J.\ Lett.}

\def\d{{\rm d}}

\begin{document}

\let\includegraphics\oldincludegraphics

\maketitle

\begin{affiliations}
 \item Charles University, Faculty of Mathematics and Physics, Institute of Astronomy, V Hole\v sovi\v ck\'ach 2, 18000 Prague 8, Czech Republic
 \item Aix Marseille University, CNRS, CNES, LAM, Institut Origines, Marseille, France
 \item European Southern Observatory (ESO), Alonso de Cordova 3107, 1900 Casilla Vitacura, Santiago, Chile
 \item Department of Earth, Atmospheric and Planetary Sciences, MIT, 77 Massachusetts Avenue, Cambridge, MA 02139, USA
 \item Department of Space Studies, Southwest Research Institute, 1050 Walnut Street, Suite 300, Boulder, CO 80302, USA
\end{affiliations}


\begin{abstract}
Understanding the origin of bright shooting stars and their meteorite samples
is among the most ancient astronomy-related questions
that at larger scales has human consequences
\cite{Chladni_1794,Biot_1803AnP....15...74B,Kulik_1922}.
As of today, only ${\sim}\,6\%$ of meteorite falls have been firmly linked to their sources 
(Moon, Mars, and asteroid (4) Vesta;
\cite{Marvin_1983GeoRL..10..775M,Treiman_2000P&SS...48.1213T,Thomas_1997Sci...277.1492T}).
Here, we show that ${\sim}\,70\%$ of meteorites originate
from three recent breakups of $D > 30\,{\rm km}$ asteroids
that occurred 5.8, 7.6 and less than ${\sim}\,40$ million years ago.
These breakups,
including the well-known Karin family \cite{Nesvorny_2002Natur.417..720N},
took place in the prominent yet old Koronis and Massalia families
and are at the origin of the dominance of H and L ordinary chondrites
among meteorite falls.
These young families distinguish themselves amidst all main belt asteroids
by having a uniquely high abundance of small fragments.
Their size-frequency distribution remains steep for a few tens of millions of years,
exceeding temporarily the production of metre-sized fragments by the largest old asteroid families
(e.g., Flora, Vesta).
Supporting evidence includes the existence of associated dust bands
\cite{Sykes_1990Icar...85..267S,Reach_1997Icar..127..461R,Nesvorny_2003ApJ...591..486N},
the cosmic-ray exposure ages of H-chondrite meteorites
\cite{Graf_1995JGR...10021247G,Eugster_2006mess.book..829E},
or the distribution of pre-atmospheric orbits of meteorites
\cite{Spurny_2020M&PS...55..376S,Brown_2016Icar..266...96B,Jenniskens_2016Icar..266..384J}.
\end{abstract}


According to both dynamical models
\cite{Bottke_2002Icar..156..399B,Granvik_2018Icar..312..181G,Nesvorny_2023AJ....166...55N}
and observational surveys
\cite{Binzel_1996Sci...273..946B,Vernazza_2008Natur.454..858V,Marsset_2022AJ....163..165M},
the majority of meteorites are thought to have their origin in the main asteroid belt.
Collisions between millions of asteroids generate fragments,
which drift through the Yarkovsky thermal effect,
until they approach mean-motion or secular resonances
and planets strongly perturb eccentricities
to the point where orbits become Earth-crossing
\cite{Farinella_1998Icar..132..378F}.
However, it is exceedingly challenging to determine the provenance of the different meteorite groups
(e.g., H, L, LL, CM)
using current telescopic and spacecraft data alone,
as plausible parent bodies or parent families are not spectrally/compositionally unique
(e.g., \cite{Bus_2002Icar..158..146B,DeMeo_2009Icar..202..160D,Vernazza_2014ApJ...791..120V,Vernazza_2016AJ....152...54V}).
(4) Vesta and its family stand out as an obvious exception,
being the only possible source of most HEDs
(howardite-eucrite-diogenite; \cite{Marvin_1983GeoRL..10..775M}).
Identifying the sources of the main meteorite groups
thus remains an unresolved problem in planetary science.
Notably, meteorite falls are dominated by two groups only (H and L chondrites)
that represent ${\sim}\,70\%$ of all falls;
they are followed at significantly less proportion by LL chondrites ($8\%$) and HEDs ($6\%$).
On the contrary, kilometre-sized asteroids in the main belt,
as well as near-Earth objects (NEOs),
typically have a different composition,
with LL-like bodies being as abundant as H- or L-like bodies
\cite{Vernazza_2008Natur.454..858V,Deleon_2010A&A...517A..23D,Vernazza_2014ApJ...791..120V}.
Specifically, the Flora (LL) and Vesta (HED) families comprise the largest numbers
of kilometre-sized asteroids among all H-, L-, LL- and HED-like families
(SI Fig.~1).
Consequently, neither prominent asteroid families
nor the background population
are likely significant sources of the meteorite flux.


Instead, a few recent stochastic collisional events may be the main source of the meteorite flux,
as suggested by the cosmic-ray exposure (CRE) ages \cite{Graf_1995JGR...10021247G}.
About 40\% of all H chondrites have young CRE ages in the 5-8\,My range,
indicating a recent breakup of an H-chondrite-like body.
The Karin family, a part of the Koronis family,
is the only known H-chondrite-like family with a formation age in the 5-8\,My range
(5.8\,My, \cite{Nesvorny_2002Natur.417..720N}).
Whereas it may explain some part of the CRE distribution it can hardly explain the older and more abundant 7-8\,My ages.

\begin{figure}
\centering
\includegraphics[height=6.0cm]{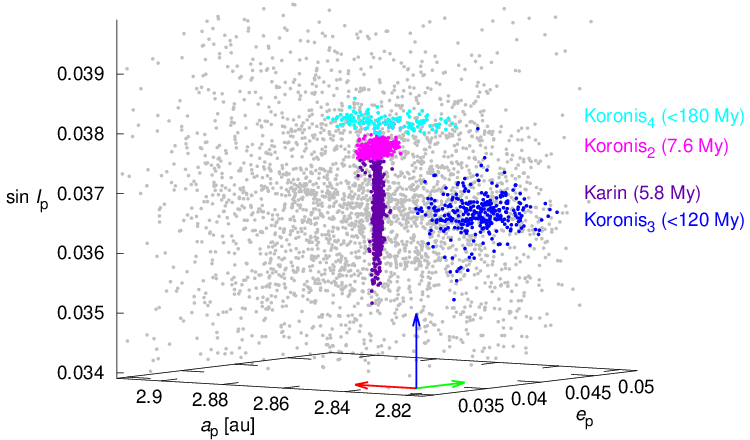}\\[.2cm]
\includegraphics[height=4.6cm]{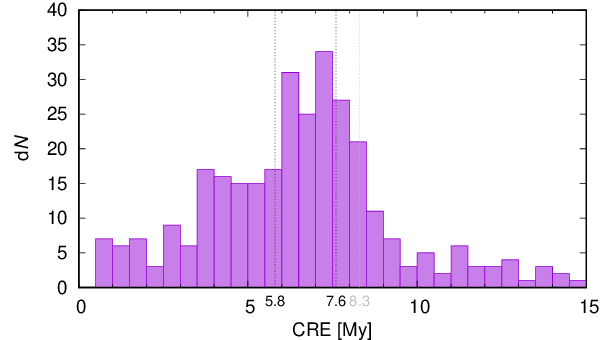}
\includegraphics[height=4.6cm]{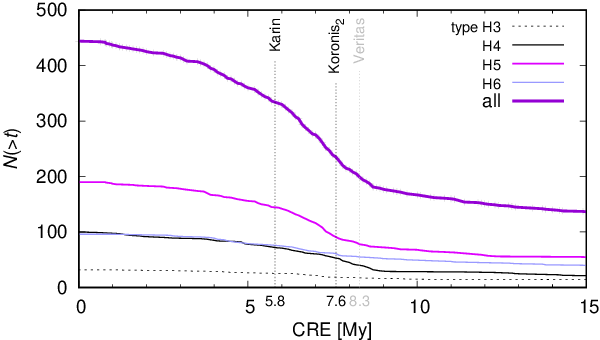}
\caption{
{\bf The Karin and Koronis$_2$ families as the main source of H chondrites.}
Top: The space of proper orbital elements
$(a_{\rm p}, e_{\rm p}, \sin I_{\rm p})$
viewed from a suitable oblique direction, when
the Karin (\color{violet}violet\color{black}) and
the Koronis$_{2}$ (\color{magenta}magenta\color{black}) families
appear as the most compact clusters.
The families have a different orientation
due to a different geometry of the breakup.
Their ages 5.8 and 7.6\,My
were determined by a convergence of orbits
(ref.~\cite{Nesvorny_2002Natur.417..720N} and this work).
Other clusters --- provisionally designated
Koronis$_{3}$ (\color{blue}blue\color{black}) and
Koronis$_{4}$ (\color{cyan}cyan\color{black})
--- are much older (possibly up to 120 and 180\,My)
and extended along the semimajor axis~$a_{\rm p}$
due to the Yarkovsky effect,
but they remained compact in the eccentricity~$e_{\rm p}$
and inclination~$\sin I_{\rm p}$.
They are no longer substantial sources of meteorites.
Bottom: The differential and the corresponding cumulative distribution
of CRE ages of H-chondrite meteorites \cite{Graf_1995JGR...10021247G},
with contributions of individual types (H3, H4, H5, H6).
Most of the meteorites exhibit ages between 5-8\,My,
which corresponds exactly to the ages of Karin and Koronis$_2$;
especially H5.
The onset at 8.3\,My is close to the age of Veritas \cite{Novakovic_2010MNRAS.402.1263N},
which may have induced a collisional cascade in the Koronis family.
}
\label{aei_KARIN1}
\end{figure}
\begin{figure}
\centering
\includegraphics[width=10cm]{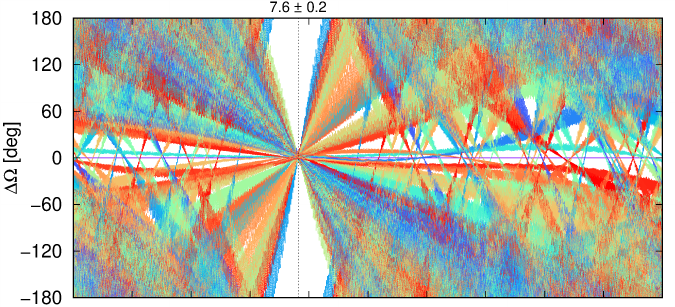}
\includegraphics[width=10cm]{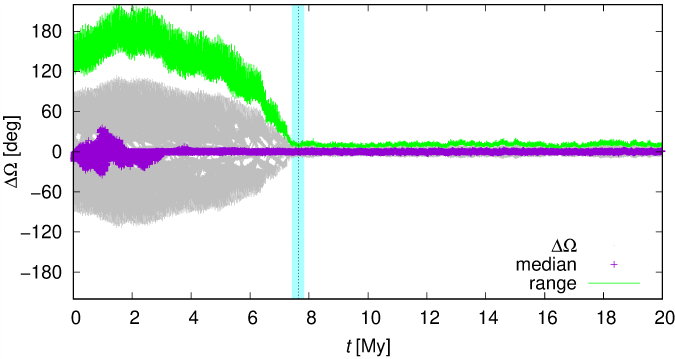}
\caption{
{\bf The Koronis$_2$ family is 7.6\,My old from convergence of orbits.}
Convergence of the longitude of nodes $\Delta\Omega$
was computed for 100 bodies and 20 clones for each body,
in order to include the Yarkovsky effect
with different rates $\dot a$ of the semimajor axis.
The rate of precession $\dot\Omega(a)$ is a non-linear function of $a$.
Top: Temporal evolution of $\Delta\Omega$ for a subset of clones (colours).
Bottom: the clones selected for each time (\color{gray}gray\color{black}),
the median (\color{violet}violet\color{black}), and
the range (\color{green}green\color{black})
of the $\Delta\Omega$ distribution.
The percentage of interlopers (which were removed) is up to 50\%,
mostly due to contamination from the neighbouring Karin family.
The orbits exhibit a convergence at $(7.6\pm 0.2)\,{\rm My}$.
}
\label{converg}
\end{figure}

To constrain the main source of H chondrites,
we searched for additional and relatively young S-type families
across the main belt
and, in particular, among all major H-like families
(Agnia, Koronis, Maria, Merxia, Phocaea).
We identified three clusters, all in the Koronis family
(Fig.~\ref{aei_KARIN1}).
Out of the three clusters, only the Koronis$_2$ family \cite{Molnar_2009DPS....41.2705M},
exhibits a convergence of orbits at the corresponding age of 
$(7.6\pm0.2)\,{\rm My}$
(Fig.~\ref{converg}; Methods).
Among the young Koronis families,
Koronis$_2$ has the steepest size-frequency distribution
(SFD; with the power-law slope $-4.0$), followed by Karin ($-2.9$).
When extrapolated to small sizes,
the SFD of Karin 'overlays' the prominent 2.1-degree IRAS dust band
\cite{Nesvorny_2006Icar..181..107N}
(Fig.~\ref{sfd_0005.800}).
A large amount of dust is released by a breakup \cite{Flynn_2009P&SS...57..119F},
but dust particles must be continuously resupplied by a collisional cascade,
because their orbits spiral inwards due to the Poynting--Robertson drag
(while their mean inclination is preserved).
This strongly supports a continuous SFD from large (sub-km) fragments,
to intermediate metre-sized bodies (i.e., precursor bodies of meteorites), and
to very small (100-$\mu{\rm m}$) dust particles.
Both Karin and Koronis$_2$ have exactly the same inclination as the 2.1-degree IRAS dust band
and it is therefore likely that the two families are at its origin.
Notably, Koronis$_2$ should dominate Karin already at sub-km sizes.
When interpolated, the two SFDs amount to a substantial number
of metre-sized bodies,
30-$60\times 10^{10}$ (Karin) and
100-200$\times 10^{10}$ (Koronis$_2$),
in the source region.


To determine whether this number of metre-sized bodies overcomes
that of the largest S-type families
(Agnia, Eunomia, Flora, Gefion, Juno, Koronis, Massalia, Maria, Merxia, Nysa, Phocaea;
SI Fig.~2),
we used a collisional model
--- specifically, a Monte-Carlo statistical approach 
(Boulder; ref.~\cite{Morbidelli_2009Icar..204..558M}; Methods) ---
to extrapolate their observed SFDs down to $D = 1\,{\rm m}$.
This extrapolation is not trivial,
because the respective slope for $D < 1\,{\rm km}$ is not constant
due to interactions with the main belt population
\cite{CampoBagatin_1994P&SS...42.1079C,OBrien_2005Icar..178..179O,Jenniskens_2016Icar..266..384J}.
For each family, the model must be set up individually,
because each of them has a different age.
Consequently, both the main belt's and the family's initial SFDs must be adapted,
so that the final SFD corresponds to the observations,
which are complete for $D \gtrsim 1\,{\rm km}$.
Every model was run at least 10 times to determine its uncertainties,
which are mostly due to the stochasticity of collisions
(see SI for more details).
Next, we used an orbital model \cite{Broz_2011MNRAS.414.2716B}
to determine the decay time scales $\tau_{\rm mb}$ of families in the main belt
and the mean lifetimes $\bar\tau_{\rm neo}$ of bodies that escaped as NEOs.
Our N-body model is based on a symplectic integrator (SWIFT; ref. \cite{Levison_Duncan_1994Icar..108...18L}).
It takes into account a number of effects driving the transport, in particular,
perturbations by 11 massive bodies (Sun, Mercury to Neptune, Ceres, Vesta),
mean-motion and secular gravitational resonances,
close encounters of NEOs and planets,
the Yarkovsky effect \cite{Yarkovsky_1901,Vokrouhlicky_1998A&A...335.1093V,Vokrouhlicky_Farinella_1999AJ....118.3049V},
the YORP effect \cite{Rubincam_2000Icar..148....2R,Capek_Vokrouhlicky_2004Icar..172..526C},
collisional reorientations, and
size-dependent spin limits
(see SI for more details).
We used approximately $10^3$ mass-less particles per family (and per size~$D$),
allowing us to estimate steady-state NEO populations as
$N_{\rm neo}({>}D) = N_{\rm mb}({>}D) \bar\tau_{\rm neo}/\tau_{\rm mb}$.


We find that the Karin and Koronis$_2$ families are far more productive in terms of meteoroids
(by at least a factor of 10)
than any of the largest families (Figs.~\ref{sfd_0005.800}, \ref{pie}).
When the Karin and Koronis$_2$ metre-sized bodies are transported from the main belt to the NEO space,
their numbers are relatively decreased due to their unfavourably short NEO lifetimes. 
Nevertheless, their abundance is still greater than the total number of metre-sized NEOs
originating from the Vesta and Flora families, in agreement with meteorite falls statistics. 


To have a better understanding of the physical process at play,
we ran our collisional evolution model with an initially steep Karin-like SFD ($-2.9$)
and let it evolve for up to 100\,My.
After 100\,My of collisional evolution (Fig.~\ref{sfd_0005.800}),
the slope of the SFD at sub-km sizes already becomes much shallower ($-1.4$)
and the number of metre-sized bodies within the family
is already less important than in the Vesta or Flora families.
This explains for example today's minimal contribution of the 100\,My-old Agnia family
to the current meteorite flux.
It follows that only recent (${\lesssim}\,40\,{\rm My}$)
yet sufficiently large ($D > 30\,{\rm km}$) breakups
can overcome the meteorite production originating
from the largest old families.


Overall, our numerical simulations produce relative abundances of H-, L-, LL- and HED-like bodies
(Fig.~\ref{pie})
that are in excellent agreement
with the compositional distribution of NEOs (within 10\%; \cite{Marsset_2022AJ....163..165M})
and also the meteorite fall statistics \cite{Gattacceca_2022M&PS...57.2102G}.
For kilometre-sized NEOs, the Phocaea, Juno and Flora families
are by far the main sources of H-, L- and LL-like NEOs, respectively.
At metre sizes, the Karin (H), Koronis$_2$ (H), Massalia$_2$ (L)
and Flora (LL) families are by far the main sources of H-, L- and LL-like meteorites.
(See ref.~\cite{Marsset_submit} for more details on Massalia.)
This is well supported by the pre-atmospheric orbits of meteorites
\cite{Spurny_2020M&PS...55..376S,Brown_2016Icar..266...96B,Meier_2023}.
As demonstrated in Extended Data Fig.~2, some H chondrites
with the semimajor axis 2.5-$2.8\,{\rm au}$ and low inclination (${\lesssim}\,3^\circ$)
directly point to the Karin and Koronis families.


There are two other major events with associated prominent dust bands,
namely ${\sim}\,40\,{\rm My}$ ago in Massalia (L-like; ref.~\cite{Marsset_submit})
and 8.3\,My ago in Veritas (CM-like;
\cite{Nesvorny_2003ApJ...591..486N,Farley_2006Natur.439..295F,Vernazza_2016AJ....152...54V}) families.
Using similar arguments as above,
they should therefore be major sources of L-like (ref.~\cite{Marsset_submit})
and also CM-like metre-sized fragments,
implying that the total meteorite flux is largely dominated
by only four recent (${\lesssim}\,40\,{\rm My}$) collisional events.
Notably, CM-like meteoroids originating from the Veritas family
should be so common (${\sim}\,3$ times more than H chondrites)
that the Earth should experience an 'extraterrestrial rain' of CM-like material
of the same order ($10^{-6}\,{\rm km}^{-2}\,{\rm y}^{-1}$)
as the total meteorite flux \cite{Chesley_2002Icar..159..423C,Nesvorny_2023AJ....166...55N}.
It follows that the bias due to atmospheric entry for the friable CM chondrites (1.5\% of the falls)
amounts to a factor of ${\sim}\,40$ with respect to the consolidated ordinary chondrites,
highlighting the critical need of sample return missions
\cite{Lauretta_2019Natur.568...55L,Watanabe_2019Sci...364..268W}
for the minute study of highly fragile extraterrestrial materials.


\let\citep=\cite
\let\citet=\cite
\let\citealt=\cite

\begin{methods}

\paragraph{Calibration of the collisional model.}

We used the collisional code called Boulder \citep{Morbidelli_2009Icar..204..558M},
which is a Monte-Carlo approach,
working with binned differential mass distributions of an arbitrary number of populations.
In our case, we used 3 populations: the main belt, one of the families and the NEO population.
The Boulder code uses a number of parameters or relations
describing how collisions between targets and projectiles produce fragments.
The principal parameter is the critical impact specific energy $Q^\star(D)$ 
(in ${\rm erg}\,{\rm g}^{-1}$), which is a function of the target size $D$.
We used the formulation of \cite{Benz_Asphaug_1999Icar..142....5B}
with modified parameters (as shown in SI Fig.~3):
\begin{equation}
Q^\star(D) = Q_0\,(D/2)^a + B\rho\,(D/2)^b\,,
\end{equation}
where
$Q_0 = 9\times 10^7$,
$a = -0.53$,
$B = 0.5$,
$b = 1.36$
(all in cgs units when applicable).
The density~$\rho$ was either $3\,{\rm g}\,{\rm cm}^{-3}$, or specific (if known precisely;
Appendix~B).
These parameters are within the range of values
tested by \cite{Bottke_2020AJ....160...14B}.
Furthermore, relations for the largest remnant mass~$M_{\rm lr}(Q)$,
the largest fragment mass~$M_{\rm lf}(Q)$,
the slope of fragment size distribution~$q(Q)$
are needed, where $Q$~denotes the impact specific energy (also in ${\rm erg}\,{\rm g}^{-1}$),
as usually scaled by $Q^\star(D)$.
For 100- and 10-km bodies, we used the relations described in
\citet{Vernazza_2018A&A...618A.154V,Sevecek_2017Icar..296..239S},
with a linear interpolation in between.
The collisional probabilities and velocities for various combinations
of populations are listed in SI Tab.~7.
Because the evolution is stochastic,
we always compute multiple (at least 10) runs
to reject rare events (e.g., Ceres catastrophic disruptions).

Our collisional model is constrained by:
  (i)~the observed main belt SFD \citep{Bottke_2015aste.book..701B},
 (ii)~the NEO SFD \citep{Harris_2015aste.book..835H},
(iii)~the Vesta family SFD,
 (iv)~Rheasylvia basin's age 1\,Gy \citep{OBrien_2014P&SS..103..131O}, and
  (v)~(4)~Vesta's cratering record \citep{Marchi_2012Sci...336..690M},
namely the heavily-cratered terrain (HCT) and 
the large diffuse craters (LDC).
The final state of the model is shown in SI Figs.~5, 6.
As mentioned above, the $Q^\star(D)$ was adjusted in order to fit the {\em tail\/}
of the observed main belt SFD. Otherwise, the synthetic populations
underestimated the observed ones (see SI Fig.~3).

We use a full transport matrix between all populations.
In fact, transport is a complex process, driven by
the Yarkovsky drift,
the YORP effect,
collisional reorientations,
spin evolution, and
gravitational resonances.
In practice, the transport from the whole main belt to the NEO space
is characterized by a size-dependent mean decay time scale~$\tau_{\rm mb}$.
The time scale of main belt bodies must be relatively long,
otherwise the NEO population is overestimated (see SI Fig.~4).
On the other hand, the transport from the NEO to a `trash bin'
is on average very short ($8\,{\rm My}$),
which is comparable to \cite{Granvik_2018Icar..312..181G}
(6~to 11\,My; see their fig.~15).

The nominal time span of our simulations is 4.4\,Gy,
to leave some space for the early evolution,
without solving a question, whether the evolution was very early or not
(cf.~\citealt{Broz_2021NatAs...5..898B}).
Of course, cratering may also be produced very early,
but hereinafter we assume no saturation and no crater erasure for simplicity.
Consequently, we should never `overshoot' the observed record.

Our modelling certainly has some caveats.
For example, the size-strength scaling law may have an additional break at dust grain sizes
\citep{Love_1993Sci...262..550L},
which would imply in an additional `wave' in the SFD;
the YORP spin-up may destroy bodies instead of affecting transport;
possibly, there are two different rheologies for S- and C-type populations;
etc.
Nevertheless, the final synthetic SFDs are {\em independent\/} on details,
because we always fit the SFDs observed {\em today\/}.


\paragraph{7.6\,My age for Koronis$_{2}$.}

In order to estimate the age of the Koronis$_{2}$ family,
we used a backward integration
and a convergence of orbits,
namely of the angles $\Omega$, or $\omega$.
Our dynamical model was simplified, but still adequate,
by assuming only 5 massive bodies (Sun and the four giant planets).
We applied a barycentric correction
and a rotation to the Laplace plane.
We used 100 orbits,
corresponding to the largest Koronis$_{2}$ family members,
with 20 clones for each of them,
sampling a uniform distribution of the obliquity ($\cos\gamma$).

The Yarkovsky effect was included,
with the thermal parameters suitable
for S-type bodies covered by regolith:
the bulk density $\rho = 2.5\,{\rm g}\,{\rm cm}^{-3}$,
the surface density $\rho_{\rm s} = 1.5\,{\rm g}\,{\rm cm}^{-3}$,
the thermal conductivity $K = 10^{-3}\,{\rm W}\,{\rm m}^{-1}\,{\rm K}^{-1}$,
the heat capacity $C = 680\,{\rm J}\,{\rm kg}^{-1}\,{\rm K}^{-1}$,
Bond albedo $A = 0.1$,
and the thermal emissivity $\epsilon = 0.9$.
Drift rates reach up to $\dot a = 0.0015\,{\rm au}\,{\rm My}^{-1}$.
The YORP effect
is not important on this time scale
(cf.~\cite{Carruba_2016AJ....151..164C}),
A collisional reorientation is also not important.

We used the symplectic integrator MVS2
from the SWIFT package \citep{Levison_Duncan_1994Icar..108...18L}.
The time step was 18.2625\,d.
We computed the mean elements \citep{Quinn_1991AJ....101.2287Q}
by sampling of the osculating ones every 1\,y,
with a sequence of filters A, A, B
and decimation factors 10, 10, 3.
Consequently, the output step was 300\,y,
in order to suppress oscillation on the orbital time scale
but not secular.
The total time span was 20\,My.

Importantly, we improved a post-processing:
(i)~for each time step,
we computed the differences of angles $\Delta\Omega$ (or $\Delta\omega$)
with respect to a reference body (e.g., (158) Koronis);
(ii)~we chose the best clone for each body;
(iii)~we sorted selected clones according to $|\Delta\Omega|$;
(iv)~we chose the percentage of bodies,
which will be discarded as interlopers,
because it is inevitable that a family contains a percentage of interlopers
(e.g., from Karin).
(v)~The result is a subset of a set of clones,
for which we compute the median and range,
because 'outliers' actually determine the age,
not 'ordinary' bodies,
which remain close to the reference body.

A verification was done on the Karin family,
with the known age of $(5.8\pm0.1)\,{\rm My}$
\citep{Nesvorny_2002Natur.417..720N}.
As possible checks, one can also assert that 
the median is close to 0,
the spins of clones are evenly distributed,
or that other angles ($\Delta\omega$) also converge.

The Koronis$_{2}$ family
exhibits a systematic convergence
for the age $7.6\,{\rm My}$
(cf. Fig.~\ref{converg}).
Its uncertainty depends on several factors.
If the interloper percentage is $50\%$,
which implies 50 converging orbits,
the range of $\Delta\Omega$ is only $7^\circ$,
and the uncertainty is only 0.2\,My.
Since a random range is $180^\circ$,
it is definitely non-random.
Regarding non-converging orbits,
about half of them converge with respect to Karin,
--- at the correct age of 5.8\,My ---
because Karin overlaps with Koronis$_{2}$
(cf. SI Fig.~17).

If the interloper percentage is decreased to ${\sim}\,25\%$,
the age would be shifted to ${\sim}9.7\,{\rm My}$,
however, given the contamination from Karin,
as well as from the old Koronis family,
we consider such solutions unreliable.
In other words, it is impossible to not remove interlopers,
if there are interlopers.
Additional systematic uncertainty of the order of 0.5\,My
is due to the Yarkovsky effect,
in particular, the uncertainty of the bulk density $\sigma_\rho$,
and to a lesser extent, by other thermal parameters.

Finally, we list 50 converging asteroids (out of 100):

\begingroup
\footnotesize
\vskip\baselineskip
\noindent
   158,
 79975,
 84465,
 87289,
 91688,
 93840,
117887,
121652,
136781,
140302,
143047,
144159,
144614,
146657,
150050,
159121,
159210,
161809,
163638,
170802,
171639,
179248,
180965,
181144,
182760,
185001,
188109,
188754,
190445,
192102,
196852,
199593,
199681,
202266,
202537,
202603,
202763,
202809,
206118,
209361,
211804,
214679,
214835,
218049,
221394,
223407,
225057,
226815,
227509,
229655.

\endgroup

\end{methods}


\begin{addendum}
\item[Data availability]
The initial conditions of simulations and data used to produce the figures
are available at \url{http://sirrah.troja.mff.cuni.cz/~mira/hchondrites/}.

\item[Code availability]
The collisional code is available at the previous URL.

\end{addendum}

\bibliography{references}

\begin{addendum}
\item[Acknowledgements]
This work has been supported
by the Czech Science Foundation through grant 21-11058S
(M.~Bro\v z and D.~Vokrouhlick\'y);
by CNES, CNRS/INSU/PNP and the Institut Origines
(P.~Vernazza).
The MIT component of this work is supported by NASA grant 80NSSC18K0849.
We thank M.~Granvik for discussions about the subject of this work.
The authors acknowledge the six referees,
including Dr. Sunao Hasegawa,
who reviewed this article and provided excellent inputs and comments.

\item[Author Contributions]
M.B. and P.V. led the research and wrote the manuscript.
M.B. computed the collisional and orbital simulations.
M.M., P.V., F.D., R.B. provided and analyzed the spectroscopic observations.
D.V., D.N. interpreted the simulations.

\item[Competing Interests]
The authors declare that they have no competing financial interests.

\item[Correspondence]
Correspondence and requests for materials should be addressed to
M.B.\hfil\break (email: mira@sirrah.troja.mff.cuni.cz).

\end{addendum}


\clearpage

\clearpage

\clearpage
\begin{figure}
\centering
\includegraphics[width=11cm]{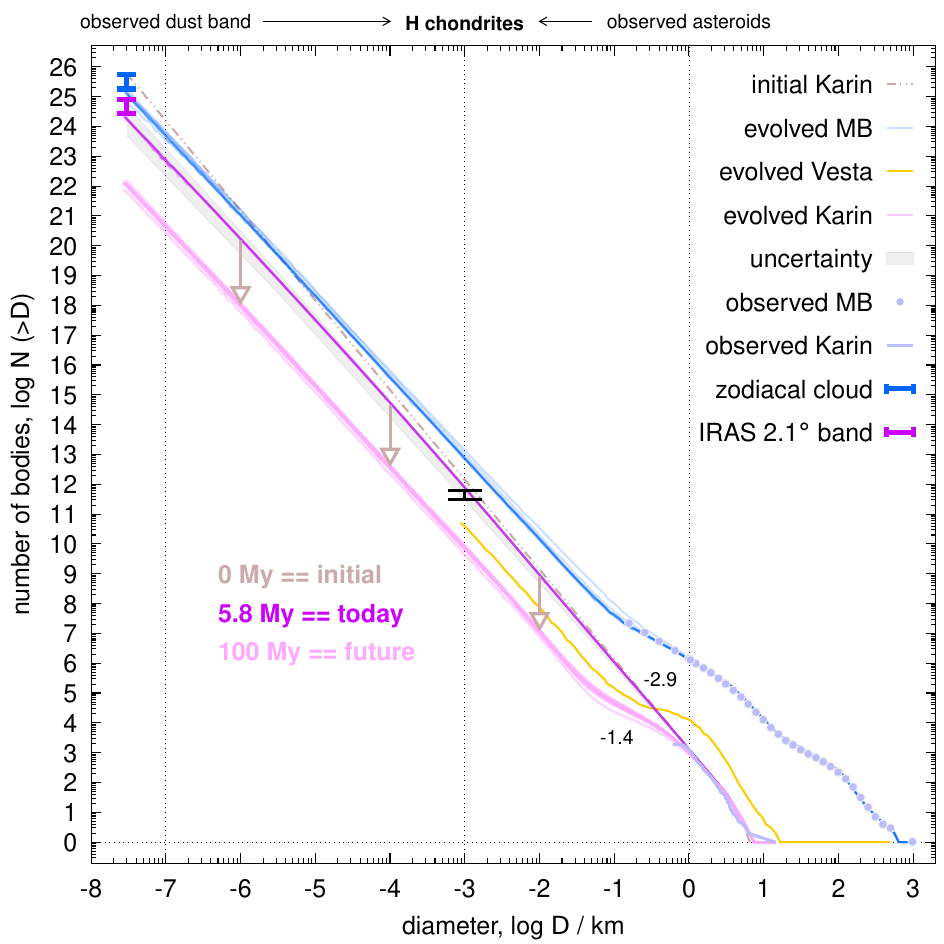}
\caption{
{\bf Excess of metre-sized bodies among young families with respect to large but old ones.}
The synthetic SFD of the Karin family was initially steep ($N({>}\,D) = CD^q$, $q = -2.9$),
i.e., close to the observed value at multi-kilometre sizes.
To create such an SFD, a ${\sim}40$-km parent body is needed.
This SFD has not evolved much over the age 5.8\,My (\color{violet}violet\color{black}).
The SFD of the Karin family observed today (\color{gray}gray\color{black})
is constrained not only at multi-kilometre sizes,
but also at ${\sim}100\,\mu{\rm m}$,
by observations of the $2.1^\circ$ IRAS dust band.
Together with the Koronis$_2$ family,
it contributes to the dust population by a~collisional cascade,
which results in a continuous SFD.
The number of dust grains is about 10 times less than in the zodiacal cloud
(without dust from Jupiter-family comets; \citealt{Nesvorny_2010ApJ...713..816N}).
The interpolated population of metre-sized H~chondrites
is indicated by an error bar.
In the future, after 100\,My of collisional evolution (\color{pink}pink\color{black}),
the SFD will become shallow ($-1.4$) at sub-km sizes
due to interactions with the main belt population
(\color{blue}blue\color{black}; cf. \cite{Bottke_2015aste.book..701B})
and the number of metre-sized bodies within young families will be lower
than in large and old families such as Vesta (\color{yellow}yellow\color{black}).
}
\label{sfd_0005.800}
\end{figure}

\clearpage
\begin{figure}
\centering
\begin{tabular}{ccc}
synthetic main belt 1-km &
synthetic NEO 1-km &
observed NEO 1-km \\
\includegraphics[width=6cm]{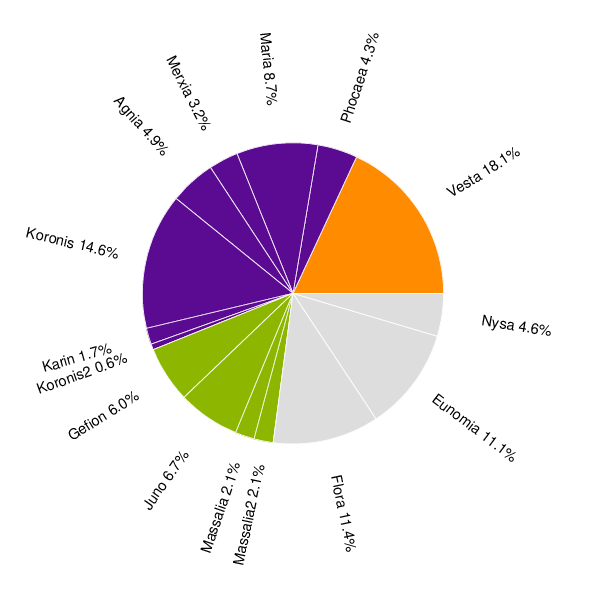} &
\includegraphics[width=6cm]{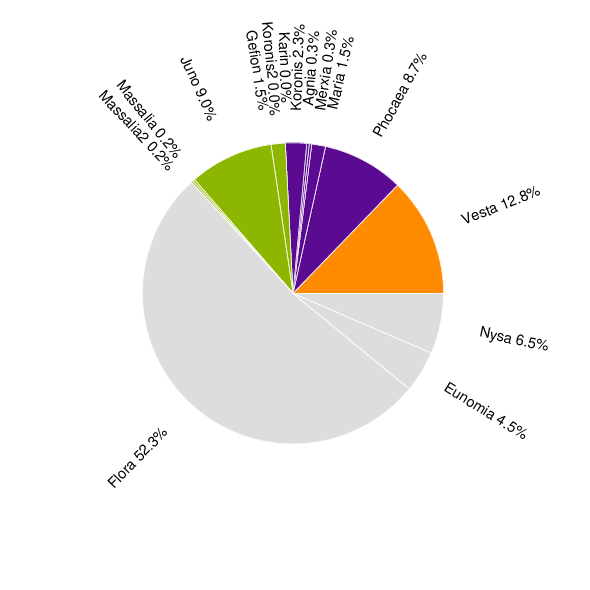} &
\includegraphics[width=6cm]{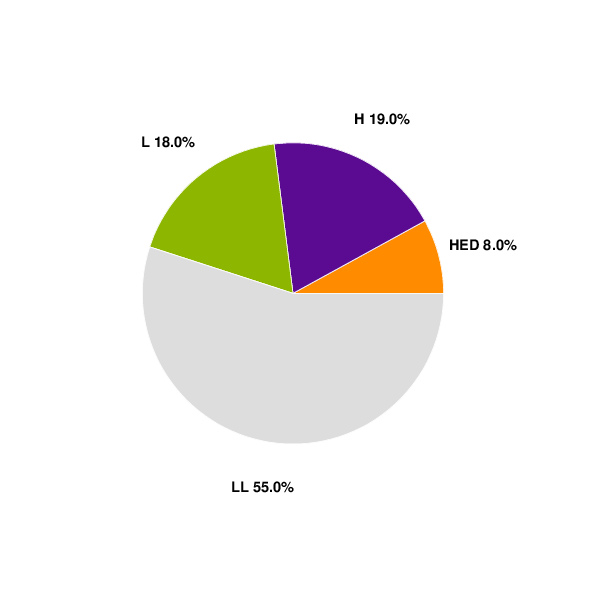} \\
synthetic main belt 1-m &
synthetic meteoroids 1-m (flux) &
observed meteoroids 1-m \\
\includegraphics[width=6cm]{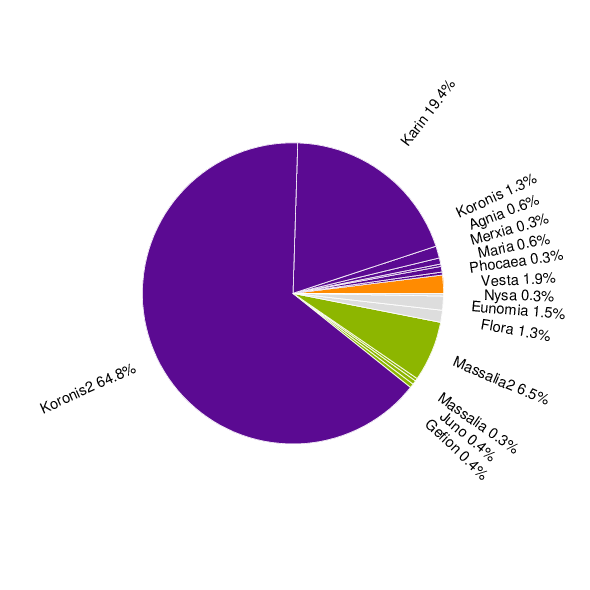} &
\includegraphics[width=6cm]{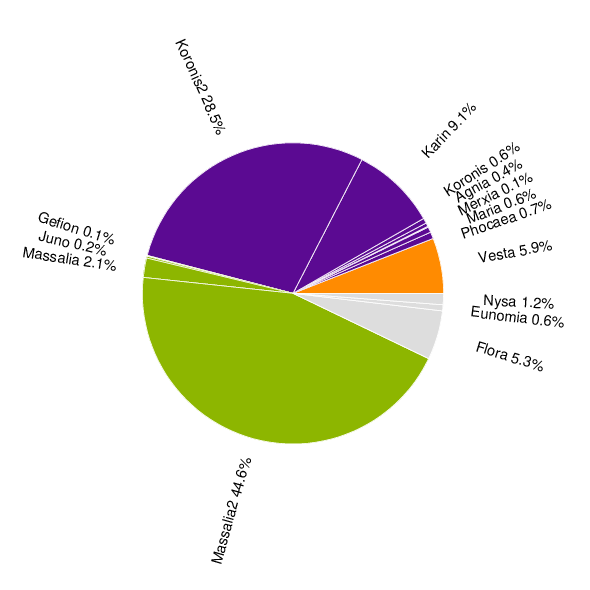} &
\includegraphics[width=6cm]{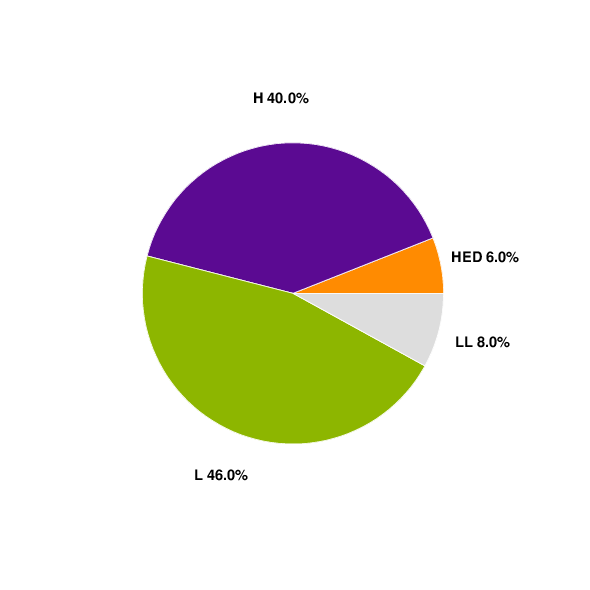} \\
\end{tabular}
\caption{
{\bf Main sources of kilometre-sized S-type NEOs and ordinary chondrite meteorites.}
Relative percentages of HED-, H-, L-, and LL-like bodies
of the synthetic main belt (left),
of the synthetic NEO (middle)
and observed NEO (right) populations are compared.
The contributions of individual families are indicated in the respective pie charts.
For 1-km NEOs, our model indicates the total percentages of
HED~12\%,
H~13\%,
L~11\%,
LL~63\%.
For 1-m meteroids,
HED~6\%,
H~40\%,
L~47\%,
LL~7\%.
Our model is in agreement
with the compositional distribution of NEOs \cite{Marsset_2022AJ....163..165M}
and the meteorite fall statistics \cite{Gattacceca_2022M&PS...57.2102G}.
}
\label{pie}
\end{figure}

\clearpage
\begin{figure}
\centering
\includegraphics[width=10.0cm]{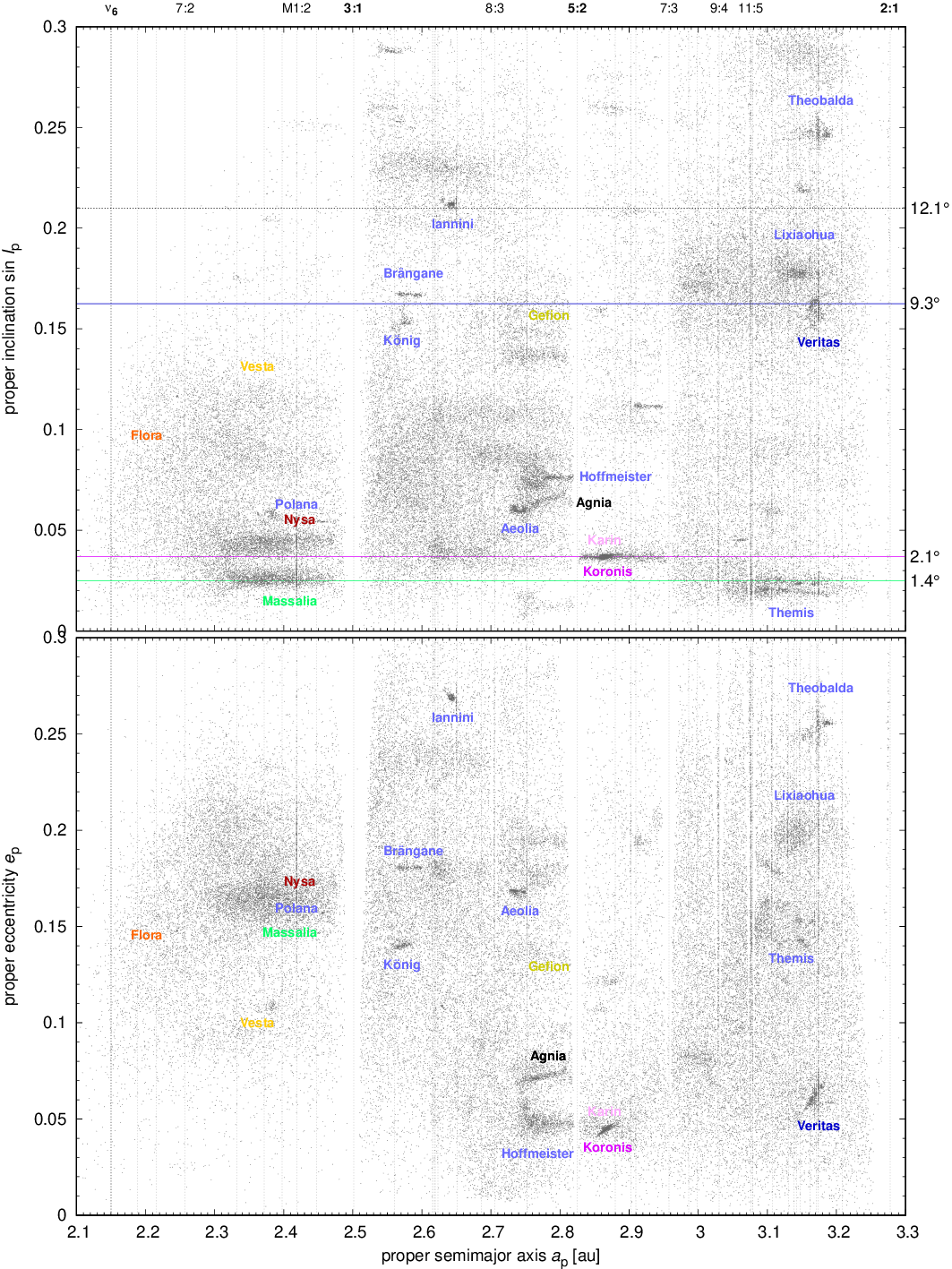}
\caption{
{\bf "Faint Main Belt", showing only bodies with the absolute magnitude
close to the limit of the Catalina Sky Survey.}
The limit has been adjusted according to the minimum distances from the Sun and the Earth, i.e.,
$H \gtrsim 19.25 + 5(\log(2.2(1-0.1)) - \log(2.2-1)) - 5(\log(a(1-e)) + \log(a-1))$,
so that for $a = 1\,{\rm au}$, $e = 0.1$, $H = 19.25\,{\rm mag}$.
The proper semimajor axis~$a_{\rm p}$
versus eccentricity~$e_{\rm p}$ (bottom)
versus inclination~$\sin I_{\rm p}$ (top) are plotted;
together with locations of the mean-motion resonances (vertical lines),
IRAS dust bands (horizontal lines),
and some of the asteroid families \citep{Nesvorny_2015aste.book..297N} (labels).
Big and old ones are almost invisible here
(e.g, Vesta, Flora, Gefion).
Small and young ones --having a steep SFD-- are prominent.
Surprisingly, the distribution of faint bodies is irregular.
The concentrations are directly related to the primary sources of meteorites:
Massalia~(L),
Karin and Koronis$_2$~(H),
Veritas~(CM),
and others.
}
\label{aei10_nature}
\end{figure}



\def\tablefoot#1{%
 \par\vspace*{2ex}%
 \parbox{\hsize}{\leftskip0pt\rightskip0pt
 {\noindent\small{\bf Notes.}~#1\par}}%
}

\clearpage
\begin{figure}
{
\centering
\includegraphics[width=11cm]{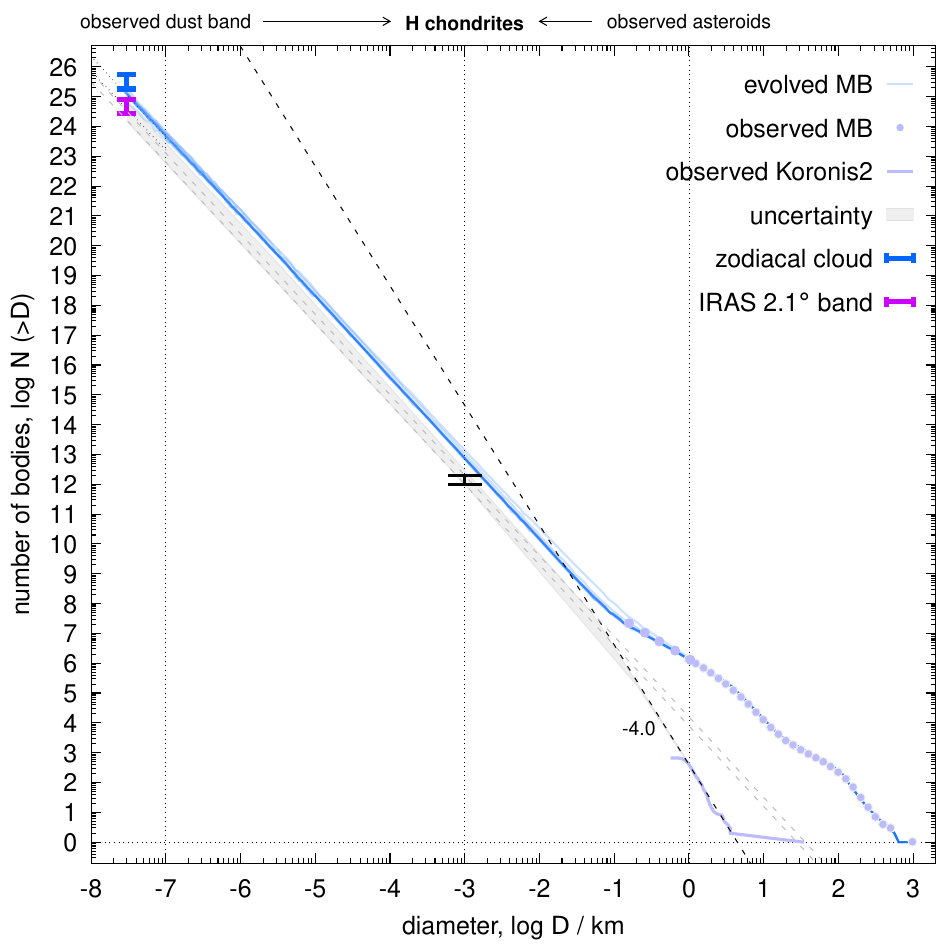}
\par
}
{\bf Extended Data Figure~1.}
{\bf Extrapolation of the observed SFD for the Koronis$_2$ family.}
The observed SFD
is extrapolated down to $0.1\,{\rm km}$
with the slope $-4.0$,
and then down to ${\sim}100\,\mu{\rm m}$
with the slope $-2.7$,
which corresponds to a collisional equilibrium.
To create such an SFD, a ${\sim}60$-km parent body is needed.
For reference, the observed $2.1^\circ$ IRAS dust band is indicated (\color{violet}violet\color{black}).
The interpolated population of metre-sized bodies is again indicated by an error bar.
\label{Koronis2_DUST}
\end{figure}

\clearpage
\begin{figure}
{
\centering
\includegraphics[width=16cm]{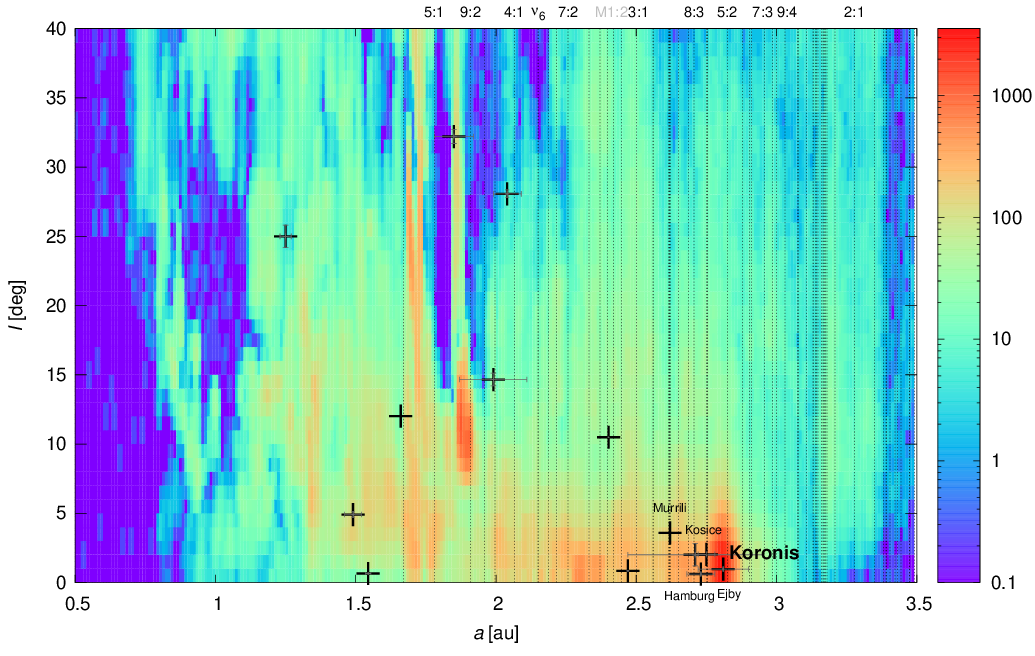}
\par
}
{\bf Extended Data Figure~2.}
{\bf Pre-atmospheric orbital elements of 14 H-chondrite falls.}
Their osculating semimajor axis versus the inclination is plotted (crosses, error bars).
A probability distribution of metre-sized meteoroids
originating from the Koronis$_2$ and Karin families
and crossing the orbit of Earth ($q < 1.3\,{\rm au}$)
is shown by colours.
Some H chondrites have the semimajor axis $2.5$-$2.8\,{\rm au}$ and low inclination (${\lesssim}\,3^\circ$),
still close to the source,
whereas other orbits have been scattered by close encounters with terrestrial planets.
Data from \cite{Meier_2023}; \url{https://www.meteoriteorbits.info/}.
\label{meteomod16_H_map}
\end{figure}

\begin{table*}
\spacing{1.1}
{\bf Extended Data Table~1.}
Dynamical time scales and cumulative numbers of 1-km asteroids
in the main belt (mb) and the near-Earth region (neo).
\par
\label{tab:1km}
{
\centering

\begin{tabular}{llrrrrrrrr}
&
&
&
1-km &
1-km &
&
1-km &
1-km &
\\
family &
res. &
$\tau_{\rm g18}$ &
$\tau_{\rm neo}$ &
$\tau_{\rm mb}$ &
$\rho$ &
$N_{\rm mb}$ &
$N_{\rm neo}$ &
obs. &
obs. \\
\vrule width 0pt depth 4pt
-- &
-- &
My &
My &
My &
${\rm g}\,{\rm cm}^{-3}$ &
$10^3$ &
1 &
1 &
\% \\
\hline
\vrule width 0pt height 9pt
Vesta (HED)         & $\nu_6$ & 6.98 & 4.39       & 1713       & 2.5  & 11.4      & 29.2       &           &          \\ 

\\                                                                                            
Phocaea (H)         & $\nu_6$ & 6.98 & 5.91       & 796        & 2.5  & 2.7       & 20.0       &           &          \\ 
Maria (H)           & 3:1     & 1.83 & 0.95$^4$   & 1533       & 3.0  & 5.5       &  3.4       &           &          \\ 
Merxia (H)          & 5:2     & 0.68 & 0.24       & 866        & 2.5  & 2.0       &  0.6       &           &          \\
Agnia (H)           & 5:2     & 0.68 & 0.19       & 1004       & 2.5  & 3.1       &  0.6       &           &          \\
Koronis (H)         & 5:2     & 0.68 & 0.82$^4$   & 1438       & 3.0  & 9.2       &  5.2       &           &          \\
Karin (H)           & 5:2     & 0.68 & $^a$       & $^a$       & 2.5  & 1.1       &  0.0       &           &          \\
\\                                                                                            
Massalia (L)        & 3:1     & 1.83 & 0.45       & 1140       & 2.5  & 2.6       &  1.1       &           &          \\ 
Gefion (L)          & 5:2     & 0.68 & 0.69       & 749        & 2.5  & 3.8       &  3.5       &           &          \\ 
Juno (L)            & 8:3     & 1.70 & 2.55       & 519        & 2.5  & 4.2       & 20.6       &           &          \\
\\
Flora (LL)          & $\nu_6$ & 6.98 & 11.93      & 722        & 2.5  & 7.2       & 119.5      &           &          \\ 
Eunomia (LL)        & 3:1     & 1.83 & 4.48       & 3078       & 3.54 & 7.0       & 10.2       &           &          \\ 
Nysa (LL)           & 3:1     & 1.83 & 4.04       & 789        & 2.5  & 2.9       & 14.8       &           &          \\
\hline                                                                                        
\vrule width 0pt height 9pt                                                                   
HED                 &         &      &            &            &      &           & 29.2       & 24        & 8        \\
H                   &         &      &            &            &      &           & 29.8       & 59        & 19       \\
L                   &         &      &            &            &      &           & 25.1       & 56        & 18       \\
LL                  &         &      &            &            &      &           & 144.5      & 172       & 55       \\
H+L+LL              &         &      &            &            &      &           & 199.4      & 287       & 92       \\
\\                                                                                            
all S-types         &         &      &            &            &      &  231 &                 & 287       &          \\
all bodies          &         &      &            &            &      & 1360 &                 & 925 \cite{Nesvorny_2023AJ....166...55N}     &    \\
\hline
\end{tabular}

\par
}
\tablefoot{
For all families, we report
the neighbouring resonances,
the NEO life time~$\tau_{\rm g18}$ from \cite{Granvik_2018Icar..312..181G},
the NEO life times~$\tau_{\rm neo}$ from this work,
computed for 1-km bodies,
the main belt life times~$\tau_{\rm mb}$,
the volumetric density of simulated bodies,
the observed cumulative number~$N_{\rm mb}({>}\,1\,{\rm km})$ of main belt bodies,
the computed cumulative number~$N_{\rm neo}$ of NEOs and meteoroids,
along with the observed $N_{\rm neo}$ from \cite{Marsset_submit},
where the original percentages were multiplied by the total number of S-type NEOs
($925\times 31\,\% \doteq 287$; \citealt{Marsset_2022AJ....163..165M}).
For comparison, the fraction of S-type main belt bodies is different
($1360\times 10^3 \times 17\,\% \doteq 231\times 10^3$; \citealt{Gradie_1989aste.conf..316G}).
Additional notes:
$^4$~4~outer planets;
$^u$~undersampled;
$^a$~after 100\,My.
}
\end{table*}

\begin{table*}
\spacing{1.1}
{\bf Extended Data Table~2.}
Same as Extended Data Tab.~1 for 1-m meteoroids.
\par
\label{tab:1m}
{
\centering
\footnotesize

\begin{tabular}{llrrrrrrrrrr}
&
&
&
1-m &
1-m &
&
1-m &
1-m &
1-m &
\\
family &
res. &
$\tau_{\rm g18}$ &
$\tau_{\rm neo}$ &
$\tau_{\rm mb}$ &
$\rho$ &
$N_{\rm mb}$ &
$N_{\rm neo}$ &
$\Phi$ &
obs. \\
\vrule width 0pt depth 4pt
-- &
-- &
My &
My &
My &
${\rm g}\,{\rm cm}^{-3}$ &
$10^{10}$ &
$10^8$ &
$10^{-9}\,{\rm km}^{-2}\,{\rm y}^{-1}$ &
\% \\
\hline
\vrule width 0pt height 9pt
Vesta (HED)       & $\nu_6$ & 6.98 & 2.50  & 115  & 2.5  & 2-7            & 4.3-15.2       & 18-62           &          \\ 
\\                                                                                           
Phocaea (H)       & $\nu_6$ & 6.98 & 7.24  & 114  & 2.5  & 0.5-1          & 3.2-6.4        & 3.2-6.5         &          \\
Maria (H)         & 3:1     & 1.83 & 1.82  &  98  & 2.5  & 0.8-2          & 1.5-3.7        & 2.3-5.7         &          \\ 
Merxia (H)        & 5:2     & 0.68 & 0.43  &  81  & 2.5  & 0.3-0.9        & 0.2-0.5        & 0.3-0.8         &          \\
Agnia (H)         & 5:2     & 0.68 & 0.34  & 103  & 2.5  & 1-2            & 0.3-0.7        & 1.6-3.7         &          \\
Koronis (H)       & 5:2     & 0.68 & 0.36  & 176  & 2.5  & 2-4            & 0.4-0.8        & 2.5-5.0         &          \\
Karin (H)         & 5:2     & 0.68 & 0.33  & 138  & 2.5  & 30-60          & 7.2-14.3       & 41-82           &          \\
\\                                                                                          
Massalia (L)      & 3:1     & 1.83 & 3.83  & 139  & 2.5  & 0.4-1          & 1.1-2.8        & 8.0-20          &          \\ 
Gefion (L)        & 5:2     & 0.68 & 0.32  &  75  & 2.5  & 0.5-1.5        & 0.2-0.6        & 0.3-0.9         &          \\ 
Juno (L)          & 8:3     & 1.70 & 1.38  & 204  & 2.5  & 0.5-1.5        & 0.3-1.0        & 0.6-1.9         &          \\
\\                                                               
Flora (LL)        & $\nu_6$ & 6.98 & 3.45  & 110  & 2.5  & 2-4            & 6.3-12.5       & 24-47           &          \\ 
Eunomia (LL)      & 3:1     & 1.83 & 1.56  & 199  & 2.5  & 1-6            & 0.8-4.7        & 1.3-7.4         &          \\ 
Nysa (LL)         & 3:1     & 1.83 & 1.79  & 114  & 2.5  & 0.5-0.8        & 0.8-1.3        & 6.3-10          &          \\
\hline                                                                                       
\vrule width 0pt height 9pt                                                                  
HED               &         &      &       &      &      &                & 4.3-15.2       & 18-62           & 6.0      \\
H                 &         &      &       &      &      &                & 12.8-26.4      & 51-104          & 33.4     \\
L                 &         &      &       &      &      &                & 1.6-4.4        & 8.9-23          & 37.8     \\
LL                &         &      &       &      &      &                & 7.9-18.5       & 31-65           & 8.2      \\
H+L+LL            &         &      &       &      &      &                & 22.3-49.3      & 91-191          & 80.9     \\
\\
all S-types       &         &      &       &      &      &                &                &                 &          \\
all bodies        &         &      &       &      &      & 400-1200       & {\bf 200-300} \cite{Harris_2021Icar..36514452H} & {\bf 740} \cite{Nesvorny_2023AJ....166...55N} &          \\
\hline
\\
with 2nd Koronis: \\
Koronis$_{2}$     & 5:2     & 0.68 & 0.33  & 138  & 2.5  & $\sim$100-200  & 23.9-47.8      & 137-274         & --       \\
H                 &         &      &       &      &      &                & 36.7-74.2      & 188-377         & 33.4     \\
\\
with 2nd Massalia: \\
Massalia$_{2}$    & 3:1     & 1.83 & 3.83  & 139  & 2.5  & $\sim$10-20    & 27.6-55.1      & 200-400         & --       \\
L                 &         &      &       &      &      &                & 29.2-59.5      & 209-423         & 37.8     \\
\end{tabular}

\par
}
\tablefoot{
$\Phi$ denotes the meteoroid flux,
dependent on the collisional probability with the Earth.
The observed percentages of meteorite falls from
\url{https://www.lpi.usra.edu/meteor/}, \url{https://metbase.org/}
are given with respect to {\em all\/} classes.
}
\end{table*}


\let\citep=\cite
\let\citet=\cite
\let\citealt=\cite

\def\tablefoot#1{%
 \par\vspace*{2ex}%
 \parbox{\hsize}{\leftskip0pt\rightskip0pt
 {\noindent\small{\bf Notes.}~#1\par}}%
}


\clearpage

\spacing{1.2}

\def\title#1{{\spacing{1.0}\large\bfseries\noindent\textsf{#1}\par}}

\title{Young asteroid families as the primary source of meteorites (SI)}

\section{Old asteroid families}\label{identification}

Before a discussion of young asteroid families,
it is necessary to `build a case' against old asteroid families.
We shall demonstrate that old ones are insufficient
to explain the origin of H- and L-chondrite meteorites
and that a contribution from young ones is inevitable.

\paragraph{Identification.}

We used recent catalogues (Jun 2021) to identify families.
We combined the following datasets:
Astorb \citep{Moskovitz_2019EPSC...13..644M},
AFP \citep{Knezevic_Milani_2003A&A...403.1165K,Novakovic_2019EPSC...13.1671N},
Wise \citep{Nugent_2015ApJ...814..117N},
Akari \citep{Usui_2011PASJ...63.1117U}, and
SDSS \citep{Parker_2008Icar..198..138P},
to obtain both orbital and physical data, whenever available.
We applied the hierarchical clustering method (HCM; \citealt{Zappala_1995Icar..116..291Z})
on proper orbital elements
with a variable cut off velocity as the initial step,
followed by an addition of halo (optional),
and a removal of interlopers.
Halo was used when a family merges with another family;
this is mitigated by using bodies brighter than a suitable magnitude limit for the HCM
and by adding fainter bodies, if their distance is smaller than another cut off velocity.
Interlopers are recognised on the basis of physical data;
unless specified otherwise, we assumed
a geometric albedo
$p_V \in (0.1; 0.5)$ and
a Sloan colour index
$a^\star \in (-0.1; 0.5)$.
Additionally, we used the relation between the absolute magnitude $H$
and the proper semimajor axis $a_{\rm p}$ \citep{Vokrouhlicky_2006Icar..182..118V}:
\begin{equation}
H(a_{\rm p}) = 5\log_{10}{|a_{\rm p}-a_{\rm c}|\over C}\,,
\end{equation}
where the parameter~$C$ determines the overall extent of the family.
Bodies are removed if $H < H(a_{\rm p})$.
The value of~$C$ is directly related to the upper limit of the age
(but {\em not\/} to the age; \citealt{Nesvorny_2015aste.book..297N}):

\begin{equation}
t_{\uparrow} = 1\,{\rm Gy}\, {C\over 10^{-4}\,{\rm au}} \left({a_{\rm c}\over 2.5\,{\rm au}}\right)^2 {\rho\over 2.5\,{\rm g}\,{\rm cm}^{-3}} \left({0.2\over p_V}\right)^{1/2}\,.
\end{equation}


\paragraph{Technical intermezzo.}

The Vesta family was associated at 100\,m/s (core) and 100\,m/s (halo).
For the first step, we used only bodies with $H \le 15\,{\rm mag}$,
for the second step $H > 15\,{\rm mag}$, so that the family is well
separated from other families.
Other parameters were:
$a_{\rm c} = 2.36151\,{\rm au}$, $C = 3.0\times 10^{-4}\,{\rm au}$,
$p_V \in (0.1; 0.7)$,
$a^\star \in (0; 0.5)$,
$i-z \in (-0.85; -0.05)$.
We considered (306) Unitas to be an interloper.

The Massalia family was associated at 30\,m/s (core) and 100\,m/s (halo);
with
$a_{\rm c} = 2.40863\,{\rm au}$, $C = 0.3\times 10^{-4}\,{\rm au}$,
$p_V \in (0.12; 0.6)$.
It was a difficult case, because it is close to the Nysa/Polana complex
and the 1:2 mean-motion resonance with Mars, which connects the two neighbouring families.

The Maria family was a simple case:
$v = 55\,{\rm m}/{\rm s}$,
$a_{\rm c} = 2.55370\,{\rm au}$, $C = 2.3\times 10^{-4}\,{\rm au}$.

The Merxia family too:
$v = 50\,{\rm m}/{\rm s}$;
with $a_{\rm c} = 2.74513\,{\rm au}$, $C = 0.5\times 10^{-4}\,{\rm au}$.

For the Agnia family, we had to choose a different central body (1020) Arcadia,
located in the densest part, not (847) Agnia itself.
The cut off velocities were 60\,m/s (core), 80\,m/s (halo);
together with $a_{\rm c} = 2.79024\,{\rm au}$, $C = 0.17\times 10^{-4}\,{\rm au}$.
The family has a structure strongly affected by the $z_1$ secular resonance,
along which the HCM associates bodies
\citep{Vokrouhlicky_2006Icar..183..349V}.

The Koronis family was associated at 55\,m/s, and
$a_{\rm c} = 2.86878\,{\rm au}$, $C = 4.3\times 10^{-4}\,{\rm au}$.
The family was extended beyond 2.96\,au,
i.e., the 7:3 mean-motion resonance with Jupiter,
which fits well within the $(a_{\rm p}, H)$ envelope.

The Gefion family was a simple case:
$v = 40\,{\rm m}/{\rm s}$,
$a_{\rm c} = 2.78381\,{\rm au}$, $C = 10^{-4}\,{\rm au}$.

The Juno family too:
$v = 40\,{\rm m}/{\rm s}$,
$a_{\rm c} = 2.66938\,{\rm au}$, $C = 10^{-4}\,{\rm au}$.

For the Flora family, we used
a~${<}15$-mag core at 110\,m/s and
a~${<}20$-mag halo at 100\,m/s.
Other parameters were
$a_{\rm c} = 2.20145\,{\rm au}$, $C = 2.1\times 10^{-4}\,{\rm au}$,
$p_V \in (0.12; 0.6)$,
$a^\star \in (0; 0.5)$,
$i-z \in (-0.3; 0.5)$.
It has a structure affected by the $\nu_6$ secular resonance.
Moreover, there is a persisting contamination from the Baptistina family.

The Eunomia family was associated at 40\,m/s;
with $a_{\rm c} = 2.64357\,{\rm au}$, $C = 2.3\times 10^{-4}\,{\rm au}$.
The (173) Ino family may be a part of Eunomia, just behind the 8:3 resonance.
Possibly, this is also the case of (53546) 2000~BY$_6$.

The Nysa family is complicated, because of several overlapping families
\citep{Walsh_2013Icar..225..283W}.
We used (135) Hertha as a central body, together with
a 15-mag core at 80\,m/s and
a 20-mag halo at 100\,m/s.
We considered both (44) Nysa, (135) Hertha to be interlopers,
given their reflectance spectra (E-, M-type).
Moreover, we suppressed the contamination from the Polana family by
$a_{\rm c} = 2.42851\,{\rm au}$, $C = 1.5\times 10^{-4}\,{\rm au}$,
$p_V \in (0.125; 0.5)$, and also
$\sin I_{\rm p} \in (0; 0.053)$.

All families as they were identified are shown in Fig.~\ref{aei2}.
In order to compute diameters from magnitudes,
we used either the measured albedos,
or the median albedo of the respective families.
The resulting SFDs are shown in Fig.~\ref{H_chondrite}.


\section{Main belt population at 1\,kilometre}

We can directly compare the main-belt populations at 1\,km,
using a straightforward extrapolation from multi-kilometre sizes,
because the recent catalogues allowed us to actually see the effect of observational bias.
The latter affects the SFDs substantially at sub-km sizes for S-type populations,
but at 1-km it can be `safely' extrapolated from multi-kilometer sizes
\citep{Hendler_2020PSJ.....1...75H}.
Approximate slopes derived for observed SFDs are listed in Tab.~\ref{tab:q}.
For H-chondrite families (see Fig.~\ref{H_chondrite}, left),
the sequence from most numerous to less numerous populations is
(in units of $10^3$ bodies):

\smallskip
\begingroup
\leftskip=\parindent\noindent
Koronis (9.2) $\rightarrow$
Maria (5.5) $\rightarrow$
Agnia (3.1) $\rightarrow$
Phocaea (2.7) $\rightarrow$
Merxia (2.0)
{\color{gray}$\rightarrow$ Karin (1.1)\color{black}};
\par
\endgroup
\smallskip

\noindent
where we also included the Karin family (to be discussed in Sec.~\ref{karin}).
For L-chondrite (middle):

\smallskip
\begingroup
\leftskip=\parindent\noindent
Juno (4.2) $\rightarrow$
Gefion (3.8) $\rightarrow$
Massalia (2.6);
\par
\endgroup
\smallskip
\noindent for LL-chondrite (right):

\smallskip
Flora (7.2) $\rightarrow$
Eunomia (7.0) $\rightarrow$
Nysa (5.7).
\par
\smallskip

On the other hand, a simple extrapolation of SFDs down to 1~metre
is not possible and we need a collisional model to do this properly.

\begin{table}
\caption{Power-law slopes of the observed SFDs of the S-type families.}
\label{tab:q}
\centering
\begin{tabular}{lrrr}
\vrule width 0pt depth 4pt
family & $q_1$ & $q_2$ & $q_3$ \\
\hline
\vrule width 0pt height 9pt
Vesta (HED)        & $-4.6$ & $-3.3$ & $-1.5$ \\
Phocaea (H)        & $-2.7$ & $-1.4$ &        \\
Maria (H)          & $-2.0$ & $-2.7$ & $-1.5$ \\
Merxia (H)         & $-3.2$ & $-2.5$ &        \\
Agnia (H)          & $-3.2$ & $-3.0$ & $-2.7$ \\
Koronis (H)        & $-2.5$ & $-1.5$ &        \\
Karin (H)          & $-4.2$ & $-2.9$ &        \\
Massalia (L)       & $-5.7$ & $-3.4$ & $-2.8$ \\
Gefion (L)         & $-3.9$ & $-1.7$ & $-1.2$ \\
Juno (L)           & $-2.8$ & $-3.7$ & $-3.1$ \\
Flora (LL)         & $-3.8$ & $-2.8$ & $-1.3$ \\
Eunomia (LL)       & $-4.5$ & $-3.2$ & $-1.2$ \\
Nysa (LL)          & $-8.9$ & $-4.3$ & $-1.7$ \\
\hline
\end{tabular}
\end{table}


\section{Main belt population at 1\,metre}\label{extrapolation}

\paragraph{Initial conditions.}

For each of the families, the collisional model must be set up individually.
The initial conditions correspond to the age of the family,
which is unknown.
Consequently, both the main belt and the family SFDs must be adapted,
so that the final conditions correspond to the observations.
The initial SFD was characterized by
the largest remnant (LR),
the largest fragment (LF), and
the power-law cumulative slopes:
$q_{\rm a}$,
$q_{\rm b}$,
$q_{\rm c}$,
$q_{\rm d}$,
with the diameter ranges specified by:
$D_1$,
$D_2$,
$D_3$.
Again, every model was run at least 10 times
to determine its uncertainties,
which are mostly due to the stochasticity of collisions,
or break-ups of large asteroids with a fractional probability.
We always tried to use the simplest initial conditions possible,
i.e., a simple power law
$q_{\rm a} = q_{\rm b}$,
which subsequently `breaks' in the course of collisional evolution,
$q_{\rm a} > q_{\rm b}$,
because it reaches equilibrium with the background population.
The values of $q_{\rm c}$ or $q_{\rm d}$ should be less steep than $-3$
to prevent a divergence of mass
(cf. Sec.~\ref{karin}).
If it did not work,
because the initial conditions were not simple,
we prepared a more complicated model(s).

Our results for relatively young families
(Merxia, Agnia, Juno, Phocaea),
as well as some old families
(Vesta, Koronis, Flora)
suggest the possibility that their SFDs were initially simple power-laws,
starting at the largest fragment and
ending even below the observational incompleteness threshold
(see Fig.~\ref{sfd_0800}).
Ages of these families are easy to estimate
(see Tab.~\ref{tab3}).
We wait until the SFD `breaks' to two power-laws
and fits the observed SFD.
The break is induced by main belt$\,\leftrightarrow\,$family or secondary collisions
and typically occurs at $D \doteq 5\,{\rm km}$.

However, the remaining families (Massalia, Maria, Gefion, Eunomia)
required more complicated initial conditions,
as demonstrated in Fig.~\ref{sfd_0800}.

In all cases, the final synthetic SFDs are almost {\em independent\/} on the ages.
This is not surprising, simply because we fit the SFDs observed {\em today\/}.
This is true also for the extrapolations to 1\,m,
because all (old) families are in a collisional equilibrium
with the main belt, already at sub- to multi-km sizes.


\paragraph{Technical intermezzo.}

Maria's synthetic SFD often `undershoots' the observed one at $D \simeq 1\,{\rm km}$
which would correspond to an age younger than 2500\,My (not to 3000\,My
as suggested by orbital models);
it is also very shallow at large sizes,
which is typical for populations of objects including interlopers.

Gefion's SFD often `overshoots' for its previously proposed age of 470\,My \citep{Nesvorny_2009Icar..200..698N}
and the only way to fit observations is again using a broken power-law.
On the other hand, if the initial SFD is a simple power-law $q_{\rm a} = q_{\rm b} = -4.6$,
the best-fit is obtained naturally for 1500\,My
which might be more compatible with
\cite{Aljbaae_2019A&A...622A..39A}.

In the case of Massalia,
a broken power-law must be used to obtain a fit at 150\,My
\citep{Vokrouhlicky_2006Icar..182..118V}.
For a simple power-law with the cumulative slopes $q_{\rm a} = q_{\rm b} = -7.5$,
the age would be as long as 800\,My
(cf. \cite{Marsset_submit}).


Eunomia's SFD at $D \simeq 20\,{\rm km}$ is wavy,
which is either related to the primordial SFD, or the presence of interlopers.
Its SFD at multi-km sizes is very shallow,
actually the most shallow of all families,
which indicates a significant depletion of objects
and a preference for an older age (definitely more than 3000\,My).

Taken overall, ages seem to be self-consistent;
none is older than 4.4\,Gy
and they are distributed over the whole interval from 0 to 4.4\,Gy.


\paragraph{Extrapolated population.}

For metre-sized bodies, there is inevitably some stochasticity,
leading to about half-order variation from simulation to simulation
in the absolute number of bodies,
due to secondary collisions and temporally variable tail.
Consequently, for H-chondrite families, the populations are
(in $10^{10}$ units):

\smallskip
\begingroup
\leftskip=\parindent\parindent=0pt
{\color{gray}Karin (30-60) $\rightarrow$\color{black}}\
Koronis (2-4) $\rightarrow$
Maria (0.8-2) $\rightarrow$
Agnia \hbox{(1-2)} $\rightarrow$
Phocaea (0.5-1) $\rightarrow$
Merxia (0.3-0.9);
\par
\endgroup
\smallskip

\noindent for L-chondrite:

\smallskip
\begingroup
\leftskip=\parindent\parindent=0pt
Juno (0.5-1.5) $\rightarrow$
Gefion (0.5-1.5) $\rightarrow$
Massalia (0.4-1);
\par
\endgroup
\smallskip

\noindent for LL-chondrite:

\smallskip
\smallskip
\begingroup
\leftskip=\parindent\parindent=0pt
Eunomia (1-6) $\rightarrow$
Flora (2-4) $\rightarrow$
Nysa (1-1.6).
\par
\endgroup
\smallskip

For Karin, see again Sec.~\ref{karin}.
Otherwise, the order is similar for metre- and kilometre-sized bodies.
Maria is similar to Agnia within stochasticity;
Juno to Gefion or Massalia;
Eunomia might be slightly more populous than Flora.
Let us recall that, at this stage, all the populations are still in the main belt;
a transport is yet to be applied.

\begin{table}
\caption{Ages of the S-type families estimated from our collisional model.}
\label{tab3}
\centering
\begin{tabular}{lr}
\vrule width 0pt depth 4pt
family & age \\
--     & My \\
\hline
\vrule width 0pt height 9pt
Vesta (HED)        & $1100\pm 100$ \\
Phocaea (H)        & $ 700\pm 100$ \\
Maria (H)          & $2500\pm 300$ \\
Merxia (H)         & $ 330\pm  50$ \\
Agnia (H)          & $ 100\pm  50$ \\
Koronis (H)        & $2200\pm 300$ \\
Massalia (L)       & $ 800\pm 100$ \\
Gefion (L)         & $1500\pm 200$ \\
Juno (L)           & $ 750\pm 100$ \\
Flora (LL)         & $1200\pm 200$ \\
Eunomia (LL)       & $4200\pm 300$ \\
Nysa (LL)          & $ 600\pm 100$ \\
\hline
\end{tabular}
\end{table}


\section{NEO population at 1\,kilometre}\label{1km}

We used an orbital model described in \cite{Broz_2011MNRAS.414.2716B}
to determine the decay time scales in the main belt
and the life times among the NEOs.
It is based on the symplectic integrator SWIFT-RMVS3 \citep{Levison_Duncan_1994Icar..108...18L}.
The dynamical model includes:
11~mutually interacting bodies (Sun, Mercury to Neptune, Ceres, Vesta),
the Yarkovsky effect \citep{Vokrouhlicky_1998A&A...335.1093V,Vokrouhlicky_Farinella_1999AJ....118.3049V},
the YORP effect \citep{Capek_Vokrouhlicky_2004Icar..172..526C},
collisional reorientations,
a mass shedding, and
the strength-dependent spin limit \citep{Holsapple_2007Icar..187..500H}.
This is supplemented by a series of digital filters
to compute mean elements \citep{Quinn_1991AJ....101.2287Q}
and proper elements \citep{Sidlichovsky_Nesvorny_1996CeMDA..65..137S}.

Some of the parameter values were common for all simulations. Namely,
a time step $\Delta t = 9.13125\,{\rm d}$,
output of osculating elements $10\,{\rm ky}$,
sampling of osculating elements $1\,{\rm y}$,
sequence of filters A, A, A, B,
decimation factors 10, 10, 10, 3,
output of mean elements $3000\,{\rm y}$,
number of samples for the Fourier transform 1024,
output of proper elements $0.1\,{\rm My}$,
a thermal capacity $C = 680\,{\rm J}\,{\rm kg}^{-1}\,{\rm K}^{-1}$,
thermal conductivity $K = 10^{-3}\,{\rm W}\,{\rm m}^{-1}\,{\rm K}^{-1}$,
thermal emissivity $\epsilon = 0.9$,
Bond albedo $A = 0.1$,
surface density $\rho = 1.5\,{\rm g}\,{\rm cm}^{-3}$,
YORP efficiency $c_{\rm YORP} = 0.33$,
reorientation time scale $B = 84.5\,{\rm ky}$, with exponents
$\beta_1 = 0.83$,
$\beta_2 = 1.33$,
and normalisations
$\omega_0 = 3.49\times 10^{-4}\,{\rm rad}\,{\rm s}^{-1}$,
$D_0 = 2.0\,{\rm m}$,
cohesive strength scale $\kappa = 2.27\times 10^7\,{\rm g}\,{\rm cm}^{-1/2}\,{\rm s}^{-2}$,
friction coefficient $s = 0.25$,
relative axial ratios
$c/a = 0.7$,
$b/a = 0.7$.

Others were specific, adapted for individual families.
We always tried to create an initial synthetic family in such a way
that -- after the long-term evolution -- it ends up as similar
to the observed family (see, e.g., \citealt{Broz_2013Icar..223..844B}).
Parameters of the principal bodies (`parent bodies') are
discussed in Appendix~\ref{app:pb}.
Probably the most important choice is the initial velocity field.
According to the rule: `either escape or not escape',
we created a distribution with the peak at about the escape speed~$v_{\rm esc}$
from the respective parent body.
For simplicity, we assumed an isotropic field
(even a cratering is approximately isotropic in shifted coordinates).
Moreover, we assumed a size-dependent relation \citep{Vokrouhlicky_2006Icar..182..118V}:
\begin{equation}
v(D) = v_5\left(D\over D_5\right)^\alpha\,.\label{v_D}
\end{equation}
The geometry in the $(a, e, \sin I)$ space is further determined
by the true anomaly~$f$ and the argument of pericentre~$\omega$.
Sometimes, these are still visible in the observed distribution of elements. This is true not only for Karin, but also for much older families
\citep{Broz_2013Icar..223..844B,Marsset_2020NatAs...4..569M}.
These parameters are listed in Tab.~\ref{tab:nbody}.

The results of our simulations are summarized in
Fig.~\ref{nbody_synthetic_decay_H},
Fig.~\ref{nbody_synthetic_nea3_H},
and the respective time scales are listed in Extended Data Tab.~1.

\begin{table}
\caption{Parameters of the synthetic families used in our orbital models.}
\centering
\begin{tabular}{lrrrrr}
family & $v_5$ & $D_5$ & $\alpha$ & $f$ & $\omega$ \\
-- & ${\rm m}\,{\rm s}^{-1}$ & km &  & deg & deg \\
\hline
\vrule width 0pt height 9pt
Vesta (HED)        & 200 & 2 & $-0.5$ &  90 & 120 \\
Phocaea (H)        &  30 & 5 & $-0.5$ &  30 &   0 \\
Maria (H)          &  50 & 5 & $-0.5$ &  90 &  90 \\
Merxia (H)         &  24 & 5 & $-0.5$ &  90 &  90 \\
Agnia (H)          &  15 & 5 & $-0.5$ &  30 &   0 \\
Koronis (H)        &  50 & 5 & $-0.5$ &  30 &  30 \\
Karin (H)          &   5 & 5 & $-0.5$ &  30 &   0 \\
Massalia (L)       &  24 & 5 & $-1.0$ &  90 & 130 \\
Gefion (L)         & 100 & 2 & $-0.5$ &  90 &  30 \\
Juno (L)           & 100 & 2 & $-0.5$ &  90 &  30 \\
Flora (LL)         & 100 & 2 & $-0.5$ &  90 &  90 \\
Eunomia (LL)       & 100?& 2?& $ 0.0$ &  90 &  50 \\
Nysa (LL)          &  35 & 5 & $-0.5$ & 135 &   0 \\
\hline
\end{tabular}
\tablefoot{
$v_5$~denotes the ejection velocity,
$D_5$~the reference size,
$\alpha$~the exponent of the distribution,
$f$~the true anomaly,
$\omega$~the argument of pericentre.
}
\label{tab:nbody}
\end{table}


\paragraph{Steady-state situation.}
To estimate the number of 1-km bodies in the NEO population,
we can assume a steady state. In this situation:
\begin{equation}
N_{\rm neo}({>}1\,{\rm km}\,|\,{\rm H}) = \int_0^\infty C N_{\rm mb}({>}1\,{\rm km})\, p({\rm H})\, {f(\tau_{\rm neo})\tau_{\rm neo}\d\tau_{\rm neo}\over\tau_{\rm mb}}\,,\label{N_neo_f}
\end{equation}
where
$C$~denotes the calibration,
$p$~the probability that the family contributes to an H-like population,
$\tau_{\rm neo}$ the life time in the NEO population,
$f$~the corresponding distribution function, and
$\tau_{\rm mb}$ the life time in the main belt population;
and similarly for 1-m size and similarly for L-like, LL-like.
For constant factors, Eq.~(\ref{N_neo_f}) simplifies to:
\begin{equation}
N_{\rm neo}({>}1\,{\rm km}\,|\,{\rm H}) = C N_{\rm mb}({>}1\,{\rm km})\, p({\rm H})\, {\bar\tau_{\rm neo}\over\tau_{\rm mb}}\,,\label{N_neo}
\end{equation}
where 
$\bar\tau_{\rm neo}$ denotes the mean lifetime in the NEO population.
Actually, this is the very reason why the median must {\em not\/} be used.
However, short-lived NEO orbits are common and long-lived ones are exceptional
(see Fig.~\ref{nbody_synthetic_nea3_H}).
In other words --- outliers determine the mean value.
One solution is to use as many orbits as possible (or orbital clones).
However, the total number of bodies entering the NEO region is limited,
because we study individual families.
In other words --- a poor sampling of $\tau$'s (hence low~$\bar\tau$)
may be more realistic than fine sampling (high~$\bar\tau$).

Moreover, the NEO orbits sometimes require a very fine time step (0.25\,d),
if the eccentricity is extreme \citep{Granvik_2018Icar..312..181G};
this problem is especially urgent for the $\nu_6$ resonance,
which pushes $e\to 1$.
For some families (Flora) we thus used $\tau_{\rm g18}$ from Extended Data Tab.~1.
Alternatively, the values of $\tau$'s differ from \citep{Granvik_2018Icar..312..181G},
because some families (Flora) were identified as dense clusters,
but they might be more extended, with bodies scattered across
the $\nu_6$ resonances.

Today, the Flora family seems to provide a dominant contribution
to the population of kilometre-sized NEOs,
followed by Vesta, Phocaea, Juno.
This approximately corresponds to the percentages of observed NEOs.
However, we should take into account also the background population
which might be substantial.
It is probably not surprising, because the 11 families discussed in this work
only contain $54.1\times10^3$ of S-type 1-km bodies out of ${\sim}\,231\times10^3$
present in the main belt, i.e., less than one fourth.
One possible interpretation is that the background population
is indeed spectrally similar to the families
(cf. the "crime scene" figure in \cite{Nesvorny_2015aste.book..297N}).


\paragraph{Non-stationary situation.}
If we relax the assumption above, we have to compute the 
dynamical decay and transport from the main belt$\,\rightarrow\,$NEO
as non-stationary:
\begin{equation}
\dot N_i = -{1\over\tau_i}N_i\,,\label{N_i}
\end{equation}
\begin{equation}
\dot N_j = +{1\over\tau_i}N_i - {1\over\tau_j}N_j\,,\label{N_j}
\end{equation}
where the index
$i = 1..M$ corresponds to the families,
$j = 1..M$ to the NEO populations, respectively.
If $\dot N_j = 0$ is assumed,
Eq.~(\ref{N_j}) simplifies to Eq.~(\ref{N_neo}).

To demonstrate how contributions change in the course of time,
due to dynamical decay alone,
we solved the set of Eqs.~(\ref{N_i}) and (\ref{N_j}),
and plotted the solution in Fig.~\ref{famdecay1_1km_exponential}.
Of course, a collisional decay occurs at the same time;
it should be solved self-consistently by a collisional model.
Nevertheless, Fig.~\ref{famdecay1_1km_exponential} suggests
that family contributions to the NEO population in the past
must have been variable.
It may also suggest a lower collisional activity between
approximately 4~and 2.5\,Gy ago,
but it sensitively depends on the individual ages of the families
(cf.~Sec.~\ref{shock}).

Given the overall decay of individual families (both collisional and orbital),
they can hardly be exactly in an steady state.
Especially for young families, Eq.~(\ref{N_neo}) might be questionable,
as so is the very method for estimation of the NEO population,
because we do not know the derivatives~$\dot N_j$'s.
In principle, we can use the observations to determine~$N_j$'s
and compute $\dot N_j$'s, but not the other way around.






\section{NEO population at 1\,metre}\label{1m}

The evolution of metre-sized bodies was computed in the same way.
Their initial conditions were modified though ---
we used the current orbits of family members,
because these fragments are continuously replenished by collisions.
The time span is relatively short, $50\,{\rm My}$,
which is sufficient to measure the decay time scale.
Our results are summarised in
Fig.~\ref{nbody_metresized_decay_H},
Fig.~\ref{nbody_metresized_nea3_H},
and in Extended Data Tab.~2.

The situation is more complex for metre-sized bodies compared to the km-size ones.
There are inevitable uncertainties stemming from a fluctuating `tail' of the SFDs
due to stochastic breakups of $D \sim 10$-$100\,{\rm m}$ bodies. 
HED and LL-chondrite-like families contribute comparably:
Vesta 4.3-$15.2\times 10^8$, Flora 6.3-$12.5\times 10^8$,
in agreement with the observations.
If the absolute number of {\em all\/} metre-sized NEOs is 200-300$\,\times 10^8$
\citep{Harris_2021Icar..36514452H},
and the percentages of meteorite classes
HED   6.0\,\%,
LL    8.2\,\%,
one would expect
12.0-18.0$\,\times 10^8$,
16.4-24.6$\,\times 10^8$,
respectively.
This is not far from our synthetic numbers, given the fact
that scattered V-types (not associated with Vesta) also contribute to HEDs
and that other families (Eunomia, Nysa) also contribute to LLs.
Moreover, we computed the flux (in $10^{-9}\,{\rm y}^{-1}\,{\rm km}^{-2}$ units):
\begin{equation}
\Phi = p N_{\rm neo}({>}\,1\,{\rm m})\,,
\end{equation}
where $p$ is the collisional probability of meteoroids with the Earth,
evaluated from our orbital simulations of metre-sized bodies
(Tab.~\ref{tabc2}).
It turns out, however, that at least for the most relevant families the fluxes
are not so different from populations;
with the obvious exception of Phocaea.
Moreover, some meteoroids might be more fragile (e.g., carbonaceous chondrites),
and preferentially disintegrate during their atmospheric entry,
which would decrease the absolute numbers above.

On the contrary, H- and L-chondrite-like bodies are underestimated
compared to the observations.
If the percentages are
H    33.4\,\%,
L    37.8\,\%,
one would expect up to
67-100$\,\times 10^8$,
76-113$\,\times 10^8$
bodies, respectively.
This is different by a factor of more than ${\sim}10$.
While this is a serious mismatch (`conundrum'),
it is a confirmation that other families,
possibly much younger, should be taken seriously.

\paragraph{Uncertainties.}

In Extended Data Tab.~2, we accounted for the major uncertainty of $N_{\rm mb}({>}1\,{\rm m})$,
which stems from a stochastic variability of the SFD tail.
We believe this is (by far) the largest uncertainty, which is typically a factor of 2.
This propagates to $N_{\rm neo}({>}1\,{\rm m})$, by means of Eq.~(\ref{N_neo}).
Even with this, none of our conclusions depends on this uncertainty.

Regarding other uncertainties,
$N_{\rm mb}({>}1\,{\rm km})$ is about 10\,\%, due to extrapolation from multi-km sizes;
$\tau_{\rm mb}(1\,{\rm km})$, about 30\,\%, due to dependency on $\rho$%
\footnote{If systematically offset for all families, relative numbers remain the same.};
$\tau_{\rm mb}(1\,{\rm m})$, {\em ditto\/}, influenced by the YORP model;
$\tau_{\rm neo}(1\,{\rm km})$, about 20\,\%, less dependent on $\rho$,
possibly stochastic (if the number of asteroids is limited);
$\tau_{\rm neo}(1\,{\rm m})$, {\em ditto\/};
$N_{\rm neo}({>}1\,{\rm km})$, cf. Eq.~(\ref{N_neo}).

\paragraph{Technical intermezzo.}

We estimated the relative uncertainties in the following way.
For $N_{\rm mb}({>}1\,{\rm km})$, from observed SFD;
if complete (constant slope) down to 1\,km,
the uncertainty is close to 0 (cf. surveys).
If extrapolated from multi-km, then several ranges for extrapolation
were considered, hence the range of $N_{\rm mb}$.

For $\tau_{\rm mb}$, sometimes, we computed multiple models,
with slightly different family identification,
initial conditions, $\rho$, time span, number of particles,
which leads to different $\tau_{\rm mb}$ values.
Based on this experience, we determined $\sigma_{\tau_{\rm mb}}$.

Moreover $\tau_{\rm mb}$ and $\tau_{\rm neo}$ are strongly correlated. For example,
if a longer ('infinite') time span is used, $\tau_{\rm mb}$ may become long,
$\tau_{\rm neo}$ also becomes long, and the ratio is more-or-less same.
This pattern was repeated often, as we updated our models.


\section{Young asteroid families}\label{karin}

\subsection{Karin family.}

To estimate a contribution of young families,
we first studied the well-known Karin family = FIN~610 \citep{Nesvorny_2015aste.book..297N},
i.e., a secondary breakup in the Koronis family (H)
with an age of 5.8\,My
\citep{Nesvorny_2004Icar..170..324N,Carruba_2016AJ....151..164C}.
It contains
$1.1\times 10^3$
kilometre-sized bodies and up to
$30$ to $60\times 10^{10}$
metre-sized bodies.
It is clearly a non-steady population.

Contrary to our expectations, the Karin family may contribute
more than any other family to the population of metre-sized bodies
if its initial SFD was a power-law with the cumulative slope $-2.9$ down to 1\,m.
Indeed, the observed SFD is a {\em perfect\/} power-law down to the observational completeness (Fig.~\ref{Karin_cascade})
and the `tail' of the SFD simply had not enough time to evolve;
it takes 30\,My to decrease below Koronis (Fig.~3).

An important question is: is there enough time to deliver bodies
to the NEO space? Yes and no. The expected Yarkovsky drift rate (without YORP) is up to
$0.0003$ and $0.06\,{\rm au}\,{\rm My}^{-1}$,
for 1-km and 1-m bodies, respectively.
The distance to the neighbouring resonance 5:2 is
$0.03$ to $0.05\,{\rm au}$.
Consequently, it would take about 100\,My, until kilometre-sized bodies are delivered,
but only a few My for metre-sized bodies,
depending on their spin axis orientations.

Alternatively, one can assume that metre-sized fragments
were ejected at significantly larger speeds, as in Eq.~(\ref{v_D}).
This would make even an early transport possible.
It is closely related to an equipartition of kinetic energy
between high-mass and low-mass fragments, as seen in some SPH simulations
of break-ups \citep{Vokrouhlicky_2021A&A...654A..75V}.
Nevertheless, most fragments colliding with the Earth today
must have been travelling in space for approximately 5.8\,My.


\subsection{Koronis$_{2}$ family.}

Moreover, according to our analysis of Koronis, there is not a single sub-family, but four.
The second one is Koronis$_{2}$ = FIN~621 \citep{Molnar_2009DPS....41.2705M},
originating from a cratering event on Koronis itself.
Its SFD is even steeper ($-4.0$; Fig.~\ref{Karin_cascade}),
so that it likely dominates Karin already at $D \lesssim 0.5\,{\rm km}$.

In addition, we discovered a third and a fourth family
when looking at the $a_{\rm p}, e_{\rm p}, \sin I_{\rm p}$ distribution from a suitable direction.
The concentration or correlation of orbits is shown in Fig.~1.
They are logically more dispersed,
as small fragments have already reached the resonances (5:2, 17:7).
It is a confirmation that such collisions are still ongoing within the parent family (i.e., Koronis$_{1}$).

In other families, like Eunomia, these sub-clusters are not seen,
which is an argument in favour of the collisional cascade
being driven by secondary collisions.
However, we should estimate it explicitly
(in the same way as in our collisional model).
A projectile of diameter~$d$ is needed to disrupt a target of diameter~$D$:
\begin{equation}
d = D\left({2Q\over v^2}\right)^{1\over 3}\,,
\end{equation}
where
$Q$~is the specific energy and
$v$~the projectile speed.
The frequency of collisions (in ${\rm y}^{-1}$) is:
\begin{equation}
f = p {D^2\over 4} f_{\rm g} N({>}D)\,N({>}d)\,,
\end{equation}
where
$p$~denotes the intrinsic collisional probability (in ${\rm km}^{-2}\,{\rm y}^{-1}$; Tab.~\ref{tabc1}),
$f_{\rm g}$~the gravitational focussing factor, and
$N$'s the respective numbers of available targets and projectiles.
For main belt--main belt collisions,
$D = 30\,{\rm km}$,
$Q = Q^\star$ (i.e., catastrophic disruptions),
$d = 3.9\,{\rm km}$,
$N({>}D) = 1330$,
$N({>}d) = 129000$,
we obtain
$f = 1.1\times 10^{-7}\,{\rm y}^{-1}$, or
$1/f = 9\,{\rm My}$.
Consequently, it is not surprising that we observe a Karin-like event.

On the other hand, Koronis--Koronis collisions
occur with much higher probabilities (Tab.~\ref{tabc1}),
lower impact speeds,
and much lower numbers of bodies;
$d = 9.1\,{\rm km}$,
$N({>}D) = 10$,
$N({>}d) = 145$,
hence
$f = 4.3\times 10^{-12}\,{\rm y}^{-1}$.
What we see in Koronis
is not a cascade of secondary collisions,
but rather a series of primary collisions.

There might still be some caveats in our estimates:
(i)~the Karin and Koronis$_{2}$ families had similar nodes
and similar precession rates,
while $p$'s were computed for a uniform distribution of nodes;
(ii)~even cratering events ($Q \ll Q^\star$) are capable of producing
numerous fragments;
(iii)~a population of sub-km asteroids may have a different spatial
distribution as well as $p$'s with respect to Koronis;
(iv)~a production of S-type metre-sized fragments might have been temporarily increased by another collisions
(e.g., with CM-type fragments from the Veritas family;
\citealt{Farley_2006Natur.439..295F}).


\section{IRAS dust bands}\label{iras}

The Karin family event produced also dust,
which was observed by IRAS
as the $2.1^\circ$ band of infrared radiation
\citep{Sykes_1990Icar...85..267S,Reach_1997Icar..127..461R,Nesvorny_2006Icar..181..107N},
i.e., at exactly the same inclination as the family.
The equivalent diameter of all dust particles is approximately
$D \simeq 11\,{\rm km}$
(\citealt{Nesvorny_2006Icar..181..107N}, cf. Tab.~\ref{tab:dust}).
According to the Long Duration Exposure Facility (LDEF; \citealt{Love_1993Sci...262..550L}),
the dominant size of dust particles is
$d = 100\,\mu{\rm m}$,
which corresponds to a number of particles
$N({>}\,100\,\mu{\rm m}) = 1.3\times 10^{24}$.

Our extrapolated SFD of the Karin family,
with the slope $-2.9$ determined for multi-kilometre asteroids,
predicts the number of particles
$N({>}\,100\,\mu{\rm m}) = 2.7\times 10^{23}$,
which is surprisingly close to the IRAS value
(see Fig.~3).
In other words, our SFD seems to be reliable over 8 orders of magnitude.

The factor of ${\sim}\,5$ difference indicates that
the SFD slope is (was) even steeper, possibly close to $-3.0$.
This is a special value, because it corresponds to
a log-uniform distribution in mass.
In math, it results from a reciprocal of a uniform random variable,
$1\over x$.
In our case, every order of magnitude in size
(10\,km--1\,km,
1\,km--100\,m,
\dots
1\,mm--100\,$\mu$m)
contains about the same amount of mass.
The equivalent diameter of all orders is only $8^{1/3} = 2$ times larger.
It is {\em not\/} divergent in mass, simply because we do not continue to~0.

For Koronis$_{2}$, a straight extrapolation to $100\,\mu{\rm m}$ is impossible,
because its slope is too steep ($-4.0$);
it cannot be kept due to very frequent collisions.
If one extrapolates the SFD just by one order of magnitude to 0.1\,km,
and assume a collisional equilibrium with the main belt ($-2.7$),
it turns out (Extended Data Fig.~2) that
Koronis$_{2}$ also contributes to the $2.1^\circ$ dust band,
but it can be hardly distinguished from Karin.

Interestingly, the inclination of (20) Massalia corresponds exactly to
another one of the dust bands, namely at $1.4^\circ$
(\citealt{Nesvorny_2006Icar..181..107N}, Fig.~\ref{aei2}).
This association is much more likely than with (656) Beagle
\citep{Nesvorny_2008ApJ...679L.143N},
because the temperature profile,
constrained by IRAS 12-, 24- and 60-$\mu{\rm m}$ band observations,
indicates hotter dust grains.
If true, the Massalia family (or its part) is younger than previously thought.
As discussed in \cite{Marsset_submit},
the Massalia family slope $-2.8$ seems to be in agreement with the dust population,
$N({>}\,100\,\mu{\rm m}) \simeq 4\times 10^{23}$
(see also their fig. 4).

\begin{table}
\caption{
Possible correspondence of dust bands
and family-formation events.
}
\label{tab:dust}
\centering
\begin{tabular}{rrll}
band & $D$ & family & \\
\vrule width 0pt depth 4pt
-- & km & -- & \\
\hline
\vrule width 0pt height 9pt
$1.4^\circ$  &  4         & Massalia$_{2}$ (L) \\
$2.1^\circ$  & 11         & Karin series (H) \\
$9.8^\circ$  & 14         & Veritas (C) \\
all          & $\sim\,$21 & asteroidal dust \\
all          & 46         & zodiacal cloud \\
\hline
\end{tabular}
\tablefoot{
$D$ denotes the equivalent diameter of all dust particles
from \cite{Nesvorny_2006Icar..181..107N};
where `asteroidal' means without Jupiter-family comets.
}
\end{table}


            






\section{Radiometric cosmic-ray exposure ages}\label{cre}

OCs have measured cosmic-ray exposure ages
\citep{Eugster_2006mess.book..829E},
which are unevenly distributed.
A correspondence with recent family-formation events
is summarized in Tab.~\ref{tab:cre}.
For the prominent H-chondrite peak between $5$-$8\,{\rm My}$,
by far the best candidates are
the Karin and Koronis$_{2}$ families,
as indicated by a convergence of orbits (Methods).

The distribution of L-chondrites is broad,
from $1$ to $50\,{\rm My}$,
with a faint, but statistically significant peak close to ${\sim}\,40\,{\rm My}$
\citep{Vokrouhlicky_2000Natur.407..606V}.
According to the dust bands (Sect.~\ref{iras}),
the only possible source seems to be the young Massalia$_{2}$ family.
Otherwise, the overall range is characteristic
for collisional or transport time scales of metre-sized bodies

Unfortunately, for LL-chondrites the statistics is slightly worse.
While the underlying distribution seems similar to L-chondrites,
there might be an additional peak at about 15\,My.

Regarding non-peak H chondrites, they likely originate from the same parent body.
The older Koronis$_3$ and Koronis$_4$ sub-families produced a peak in the past
(similar to the one at $5$-$8\,{\rm My}$),
which has evolved in the course of time (like the one of L chondrites).
While these two families are too old to be substantial sources of meteorites,
other Koronis family members could have been bombarded by ejecta from Karin and Koronis$_2$,
releasing surface regolith and refreshing their surfaces.
If so, the average colour of the Koronis family should be less red
than that of similarly old (${\gtrsim}\,2000\,{\rm My}$) families,
such as Maria.
It is exactly the case \citep{Nesvorny_2005Icar..173..132N}.
After correction of the spectral slope bias due to composition \citep{Vernazza_2009Natur.458..993V},
the Koronis family appears to have the same slope as relatively young (${\sim}\,1000\,{\rm My}$) families,
such as Flora.
Within Koronis, the least red family members are located close to the Karin and Koronis$_2$ families
(Fig.~\ref{vernazza_slope2_ae}).
Ejecta from their surfaces
(${\sim}\,1\,{\rm m}$ depth, i.e., ${\sim}\,10^{-4}$ volume)
should have substantially longer CRE ages.

Finally, although the statistics are still low, pre-atmospheric orbits of H chondrites with determined CRE ages
support a common origin of peak and non-peak H chondrites.
Among the four H chondrite falls that point directly to Koronis
(Ko\v sice, Murrili, Hamburg, Ejby; see Extended Data Fig.~2),
two fall within the $5$-$8\,{\rm My}$ peak (Ko\v sice and Murrili) and
two have longer CRE ages (Hamburg and Ejby).

\begin{table}
\caption{
Correspondence of cosmic-ray exposure ages of OCs
and recent family-formation events.
}
\label{tab:cre}
\centering
\begin{tabular}{rll}
exposure & family & \\
\vrule width 0pt depth 4pt
My & -- & \\
\hline
\vrule width 0pt height 9pt
         5  & Karin (H) \\
         8  & Koronis$_{2}$ (H) \\
\\
$\sim\,$40  & Massalia$_{2}$ (L) \\
\\
        15? & ? (LL) \\
  \hline
\end{tabular}
\end{table}


\clearpage

\let\oldthebibliography=\thebibliography
\let\oldendthebibliography=\endthebibliography
\renewenvironment{thebibliography}[1]{%
  \oldthebibliography{#1}%
  \setcounter{enumiv}{64}%
}{\oldendthebibliography}



\clearpage

\begin{figure}
\centering
\begin{tabular}{l}
\includegraphics[height=7.5cm]{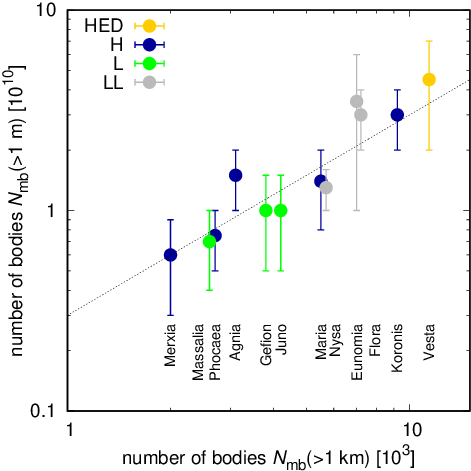} \\
\includegraphics[height=7.5cm]{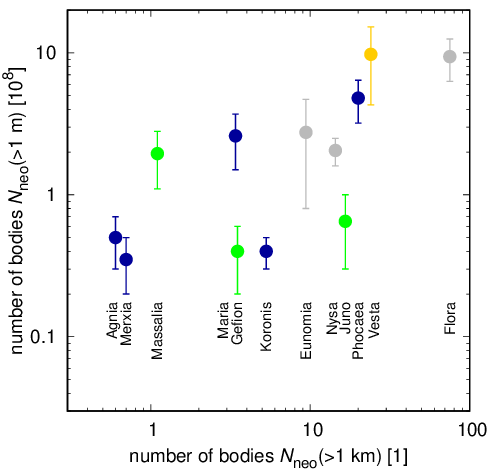}
\end{tabular}
\caption{
Illustration of the so-called 'NEO--meteorite conundrum'
\cite{Vernazza_2008Natur.454..858V}.
The Flora family, identified via previous surveys as the main source of kilometre-sized NEOs,
which we confirm here,
should also be the most productive in terms of meteoroids
along with the Vesta family.
Yet, it is not what meteorite fall statistics tell us.
Top: The numbers of bodies $N_{\rm mb}({>}\,1\,{\rm km})$ and $N_{\rm mb}({>}\,1\,{\rm m})$
in the main belt, which exhibits a positive correlation.
Bottom: The same for the NEO population originating from these families.
Individual families have been compositionally linked to meteorite classes
(H, L, LL; indicated by colours).
The number of H- or L-chondrites never exceeds that of LL-chondrites,
which is in contradiction with meteorite fall statistics
(H~40\%,
L~46\%,
LL~8\%;
\cite{Gattacceca_2022M&PS...57.2102G}).
}
\label{slope11}
\end{figure}

\begin{figure*}
\centering
\begin{tabular}{c@{\kern0.1cm}c@{\kern0.1cm}c}
\kern1cm H-chondrite &
\kern1cm L-chondrite &
\kern1cm LL-chondrite \\
\includegraphics[width=6cm]{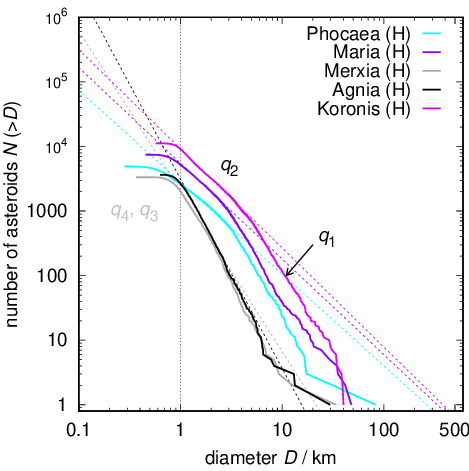} &
\includegraphics[width=6cm]{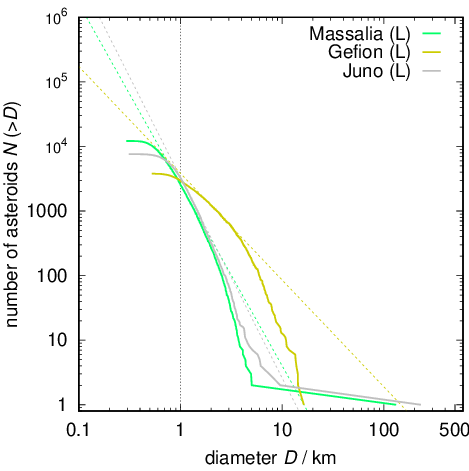} &
\includegraphics[width=6cm]{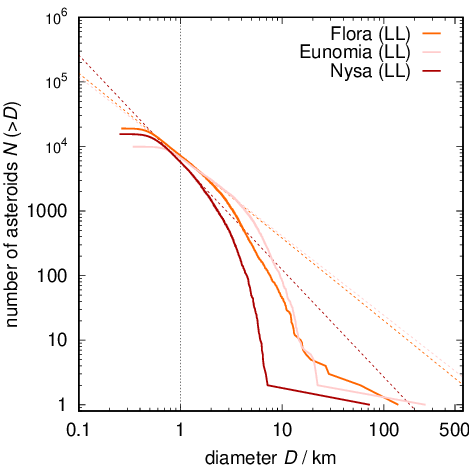} \\
\end{tabular}
\caption{
Observed cumulative size-frequency distributions (SFDs) of the S-type asteroid families:
H-chondrite-like (left),
L-chondrite-like (middle), and
LL-chondrite-like (right).
Each group is dominated by one or two families,
but it sensitively depends on the respective diameter~$D$.
For reference, $D = 1\,{\rm km}$ is indicated (black dotted line).
The SFDs exhibit the following features:
largest remnant (LR),
possibly an intermediate-size fragment,
largest fragment (LF),
first slope ($q_1$), which is steep, starting at the LF,
second slope ($q_2$), which is shallow, related to long-term collisional evolution,
third slope ($q_3$), which is even shallower, related to the scaling law
and observed break in the main belt SFD,
fourth slope ($q_4$) or bend-off, related to observational incompleteness.
}
\label{H_chondrite}
\end{figure*}

\begin{figure}
\centering
\includegraphics[width=7cm]{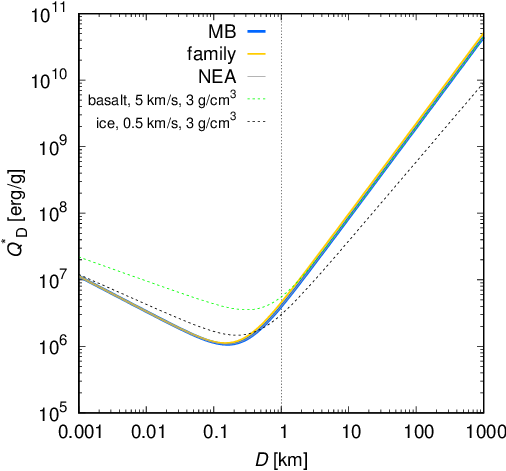}
\caption{
Modified scaling law $Q^\star(D)$ used in our collisional model.
A comparison to the nominal scaling law for basalt at 5\,km/s
(\color{green}green\color{black}\ line)
from \cite{Benz_Asphaug_1999Icar..142....5B} is also plotted.
This modification is needed to fit the main belt SFD at sub-km sizes.
}
\label{Q}
\end{figure}

\begin{figure}
\centering
\includegraphics[width=7cm]{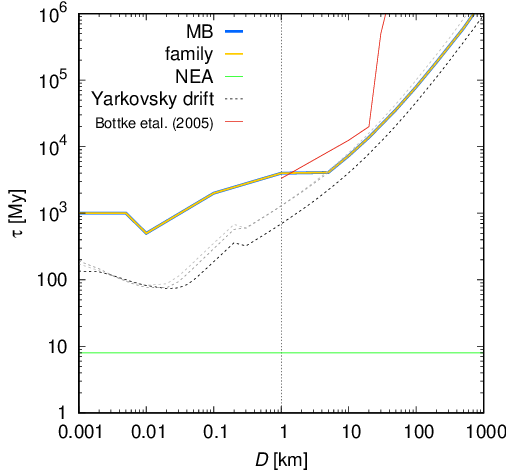}
\caption{
Dynamical decay time scales $\tau(D)$ used in our collisional model.
The main belt and families have relatively long time scales,
which are needed to fit the NEO population,
being transported from the main belt and
having a short time scale ($8\,{\rm My}$).
A comparison to the nominal time scale (\color{red}red\color{black}\ line)
of \cite{Bottke_2005Icar..179...63B} is also plotted.
For the Yarkovsky drift without spin axis evolution,
decay would be significantly shorter.
}
\label{yarko}
\end{figure}

\begin{figure}
\centering
\begin{tabular}{c}
\kern0.5cm Vesta (HED), 1100\,My \\
\includegraphics[width=7cm]{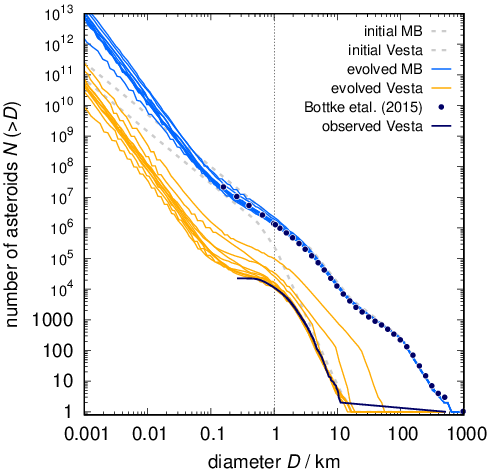} \\
\end{tabular}
\caption{
Synthetic SFD of the main belt and the Vesta family, used for calibration.
The age $1100\pm100\,{\rm My}$ is consistent with \cite{OBrien_2014P&SS..103..131O}.
}
\label{sfd_1100}
\end{figure}

\begin{figure}
\centering
\begin{tabular}{c}
\kern0.5cm NEOs, steady-state \\
\includegraphics[width=7cm]{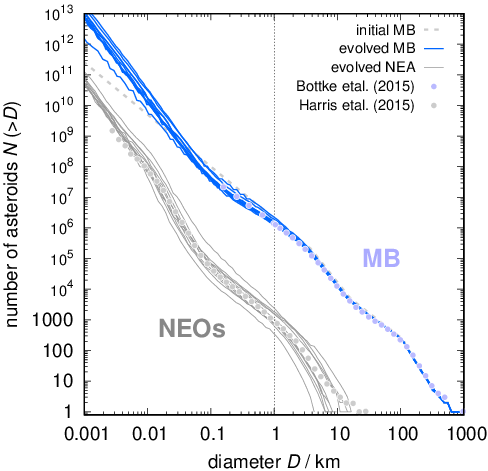} \\
\end{tabular}
\caption{
Same as Fig.~\ref{sfd_1100} for the NEOs.
}
\label{sfd_1100_NEA}
\end{figure}

\begin{figure*}
\centering
\begin{tabular}{c@{\kern0.1cm}c@{\kern0.1cm}c}
\kern0.5cm Phocaea (H), 700\,My &
\kern0.5cm Maria (H), 2500\,My &
\kern0.5cm Merxia (H), 330\,My \\
\includegraphics[width=6.0cm]{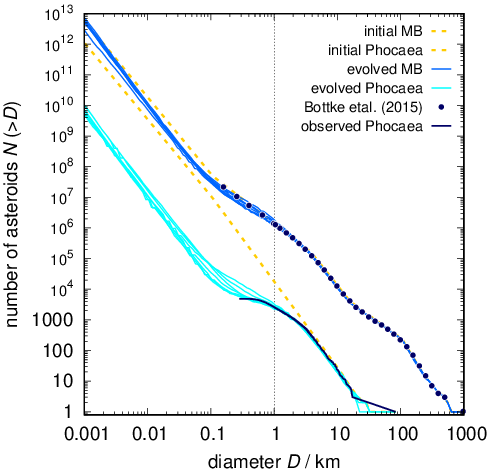} &
\includegraphics[width=6.0cm]{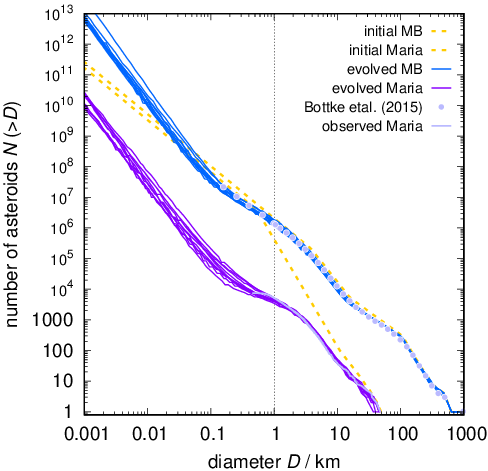} &
\includegraphics[width=6.0cm]{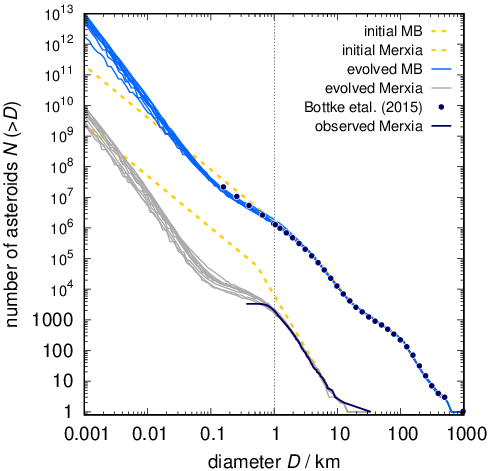} \\[0.1cm]
\kern0.5cm Agnia (H), 100\,My &
\kern0.5cm Koronis (H), 2500\,My &
\kern0.5cm Massalia (L), 800\,My \\
\includegraphics[width=6.0cm]{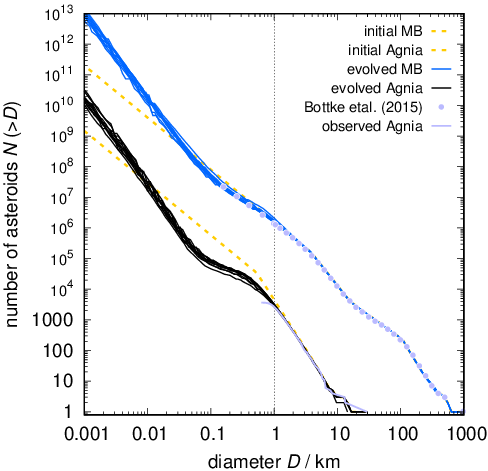} &
\includegraphics[width=6.0cm]{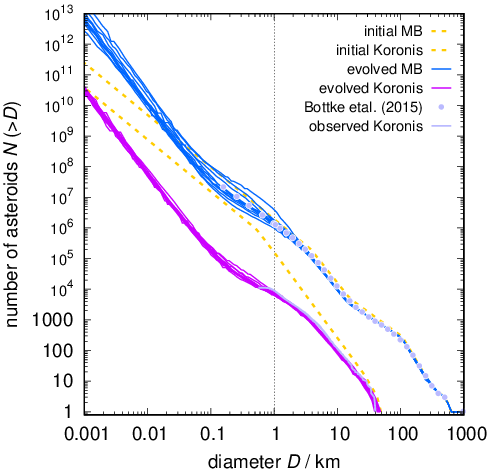} &
\includegraphics[width=6.0cm]{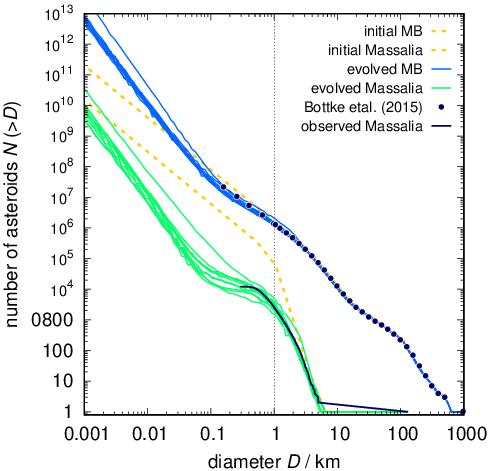} \\[0.1cm]
\end{tabular}
\caption{
Synthetic SFDs of the S-type asteroid families derived from our collisional model.
Every panel contains:
the initial main belt,
the initial family (yellow dotted),
evolved main belt (blue),
evolved family (different colours),
observed main belt \citep{Bottke_2015aste.book..701B},
observed family (gray solid).
The SFDs between 1 and 10\,km were initially a smooth power-law.
They evolved due to collisions
and exhibit a characteristic slope change at about 5\,km,
which is observed (see Tab.~\ref{tab:q}).
Every model was run 10 times to account for stochasticity.
The best-fit age is reported on top (see Tab.~\ref{tab3}).
}
\label{sfd_0800}
\end{figure*}

\addtocounter{figure}{-1}
\begin{figure*}
\centering
\begin{tabular}{c@{\kern0.1cm}c@{\kern0.1cm}c}
\kern0.5cm Gefion (L), 1500\,My &
\kern0.5cm Juno (L), 750\,My &
\kern0.5cm Flora (LL), 1200\,My \\
\includegraphics[width=6.0cm]{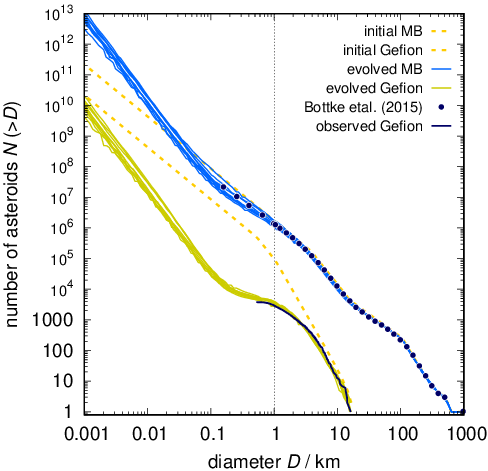} &
\includegraphics[width=6.0cm]{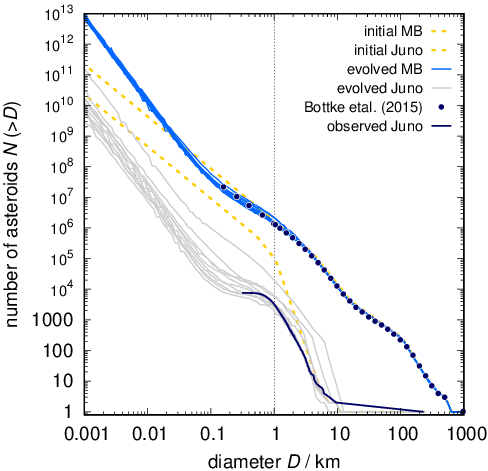} &
\includegraphics[width=6.0cm]{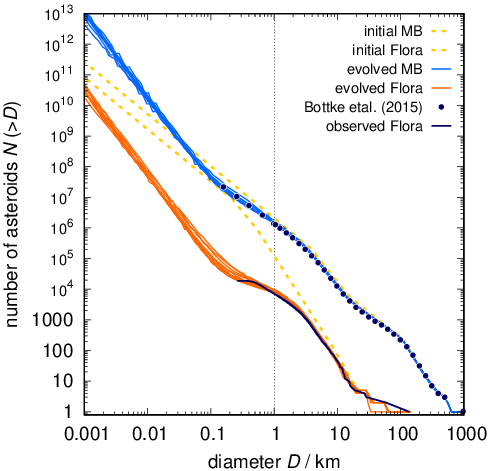} \\[0.1cm]
\kern0.5cm Eunomia (LL), 4200\,My &
\kern0.5cm Nysa (LL), 600\,My \\
\includegraphics[width=6.0cm]{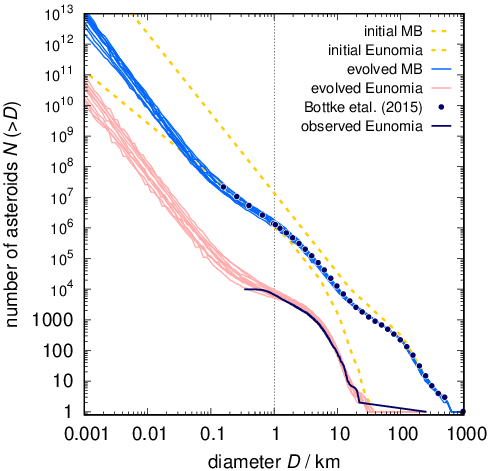} &
\includegraphics[width=6.0cm]{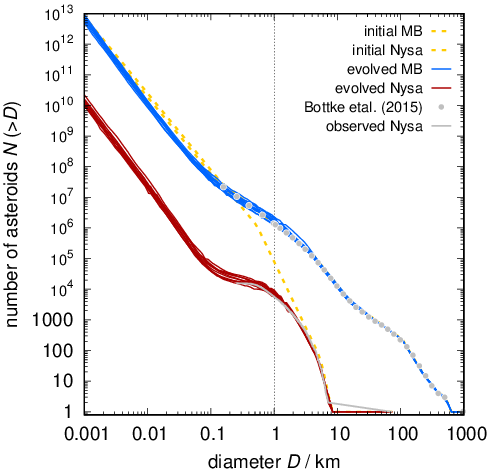} \\[0.1cm]
\end{tabular}
\caption{continued.}
\label{sfd_4200}
\end{figure*}

\begin{figure*}
\centering
\begin{tabular}{c@{}c@{}c}
\kern0.8cm H-chondrite &
\kern0.8cm L-chondrite &
\kern0.8cm LL-chondrite \\
\includegraphics[width=6.0cm]{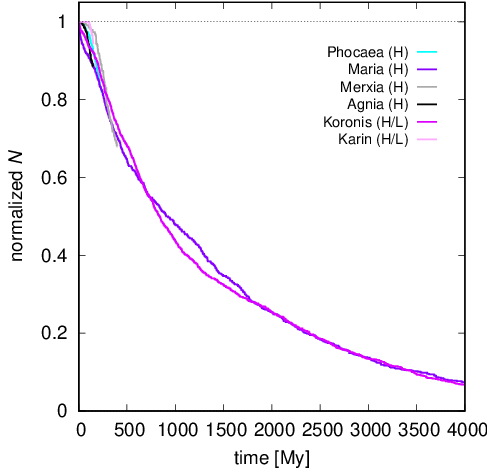} &
\includegraphics[width=6.0cm]{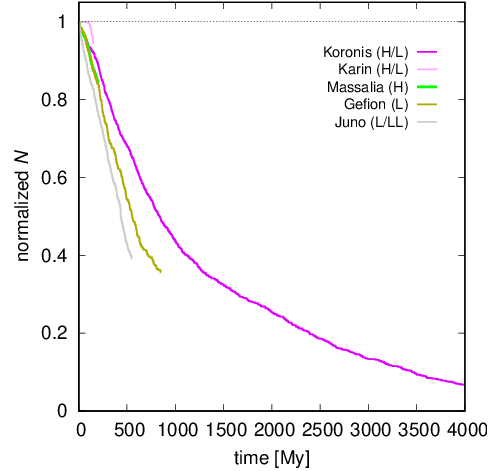} &
\includegraphics[width=6.0cm]{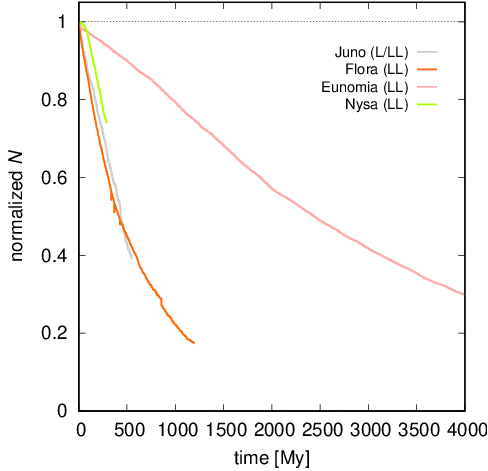} \\
\end{tabular}
\caption{
Dynamical decay of selected synthetic asteroid families derived from our orbital model:
H-chondrite-like (left),
L-chondrite (middle),
LL-chondrite (right).
Normalized number of bodies vs. time is plotted.
The decay is induced by
gravitational resonances,
the Yarkovsky drift,
as modified by the YORP effect,
collisional reorientations,
and limited by the critical frequency.
Sizes of bodies correspond to the observed SFDs;
most of them are kilometre-sized.
}
\label{nbody_synthetic_decay_H}
\end{figure*}

\begin{figure*}
\centering
\begin{tabular}{c@{}c@{}c}
\kern0.8cm H-chondrite &
\kern0.8cm L-chondrite &
\kern0.8cm LL-chondrite \\
\includegraphics[width=6.0cm]{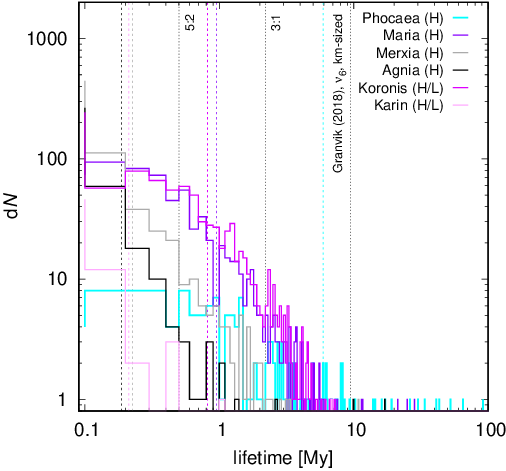} &
\includegraphics[width=6.0cm]{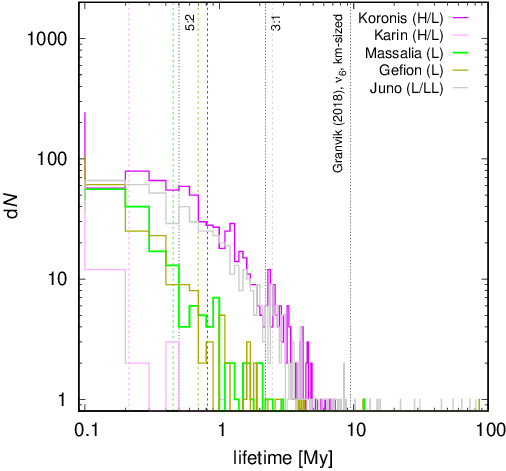} &
\includegraphics[width=6.0cm]{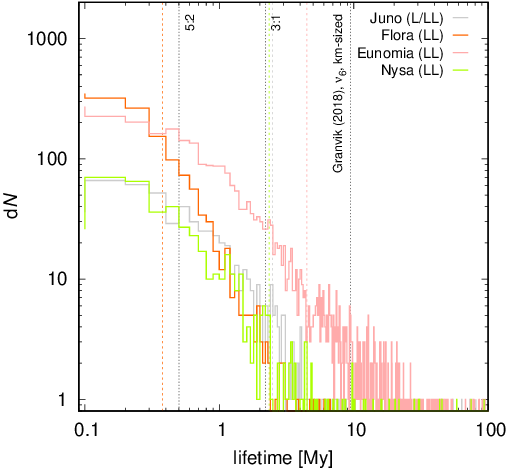} \\
\end{tabular}
\caption{
Lifetimes of bodies in the NEO space derived from our orbital model:
H-chondrite-like (left),
L-chondrite (middle),
LL-chondrite (right).
When bodies escape from the respective families via resonances
(cf. Fig.~\ref{aei2}),
they temporarily enter the NEO space.
Their lifetimes are different for different resonances,
where low-order or outer-main-belt ones tend to produce short-lived orbits,
and {\em vice versa\/}.
The mean ({\em not\/} median) lifetimes are plotted for each family (colour dashed).
For comparison, the lifetimes from \cite{Granvik_2018Icar..312..181G}
(9.4, 2.2, 0.5\,My for the $\nu_6$, 3:1, 5:2 resonances; see their Tab.~3)
are also plotted (black dotted).
}
\label{nbody_synthetic_nea3_H}
\end{figure*}

\begin{figure*}
\centering
\begin{tabular}{c@{}c@{}c}
\kern0.8cm H-chondrite &
\kern0.8cm L-chondrite &
\kern0.8cm LL-chondrite \\
\includegraphics[width=6.0cm]{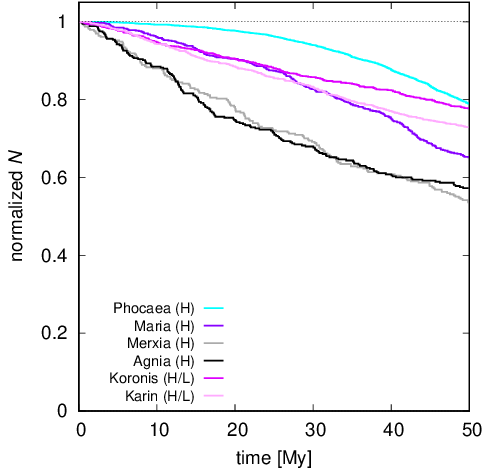} &
\includegraphics[width=6.0cm]{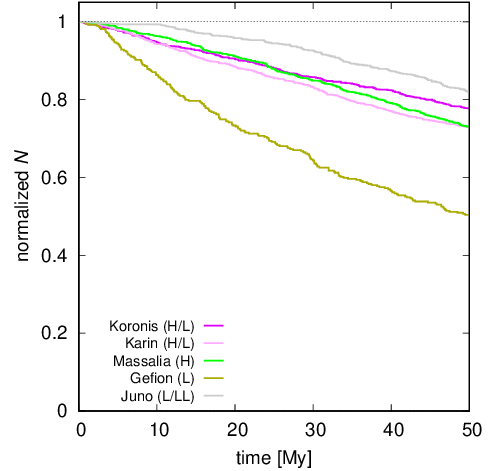} &
\includegraphics[width=6.0cm]{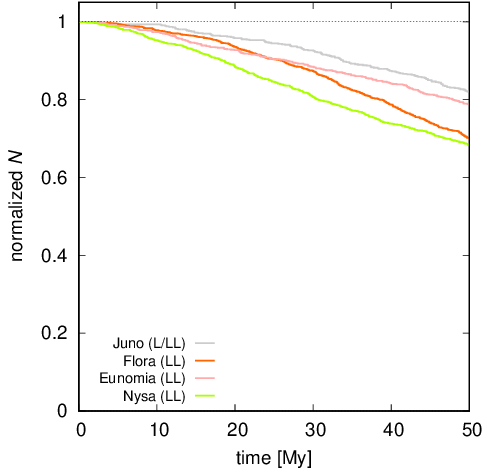} \\
\end{tabular}
\caption{
Same as Fig.~\ref{nbody_synthetic_decay_H}, but for metre-sized bodies.
}
\label{nbody_metresized_decay_H}
\end{figure*}

\begin{figure*}
\centering
\begin{tabular}{c@{}c@{}c}
\kern0.8cm H-chondrite &
\kern0.8cm L-chondrite &
\kern0.8cm LL-chondrite \\
\includegraphics[width=6.0cm]{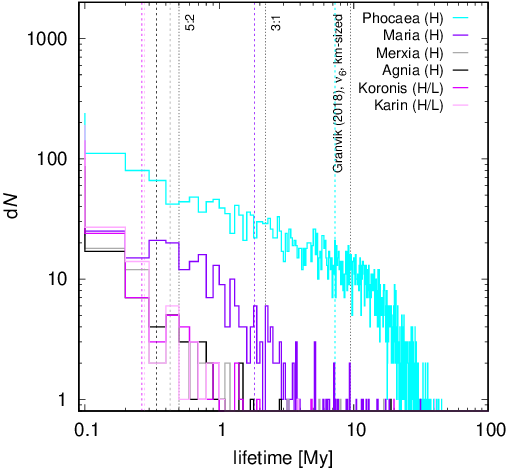} &
\includegraphics[width=6.0cm]{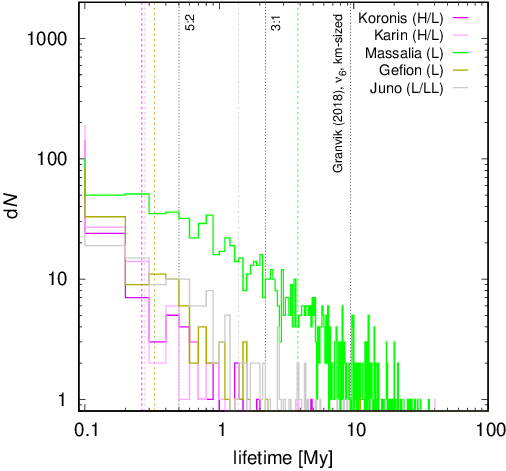} &
\includegraphics[width=6.0cm]{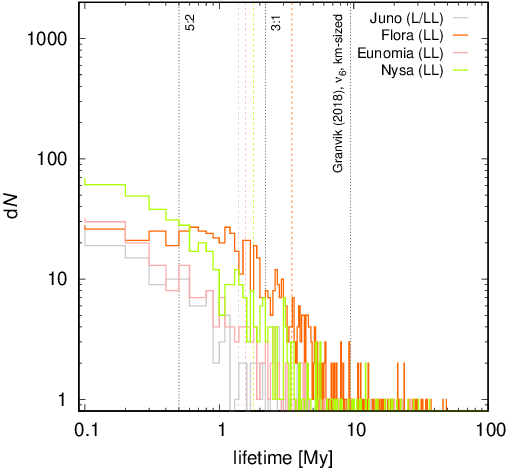} \\
\end{tabular}
\caption{
Same as Fig.~\ref{nbody_synthetic_nea3_H}, but for metre-sized bodies.
}
\label{nbody_metresized_nea3_H}
\end{figure*}

\begin{figure}
\centering
\includegraphics[width=9cm]{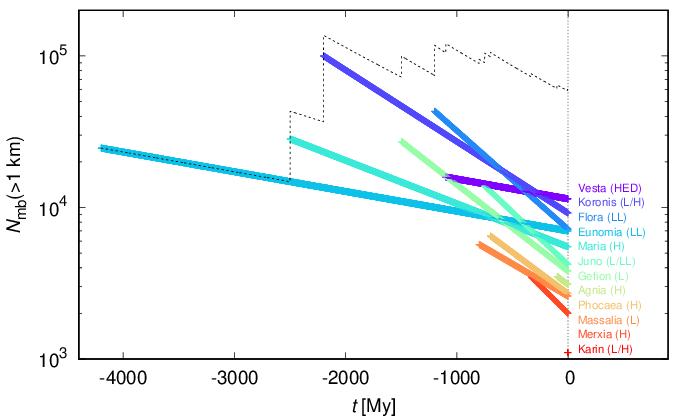}
\caption{
Extrapolated contributions of asteroid families to the population
of kilometre-sized bodies in the main belt.
The observed number $N_{\rm mb}({>}\,1\,{\rm km})$ is on the right ($t = 0$).
Here we account for the dynamical decay only
(see Extended Data Tab.~1; column~$\tau_{\rm mb}$),
so that at the family origin ($t = t_0$) the population was large
and decayed as $\exp(-(t-t_0)/\tau_{\rm mb})$.
The total of all selected families is indicated (dashed line).
The total of all main belt bodies is $1.36\times10^6$.
}
\label{famdecay1_1km_exponential}
\end{figure}

\begin{figure}
\centering
\includegraphics[width=8cm]{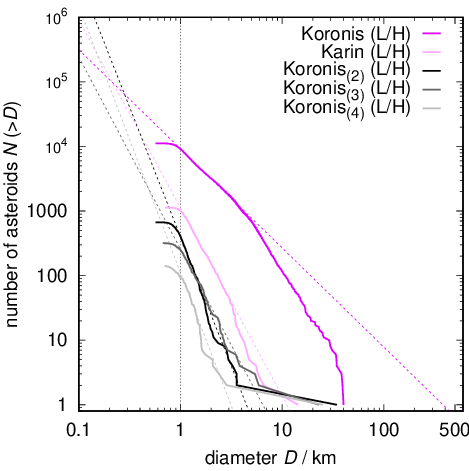} \\
\caption{Same as Fig.~\ref{H_chondrite} for the Karin, Koronis$_{2}$, Koronis$_{3}$, Koronis$_{4}$ families.}
\label{Karin_cascade}
\end{figure}

\begin{figure}
\centering
\includegraphics[width=8cm]{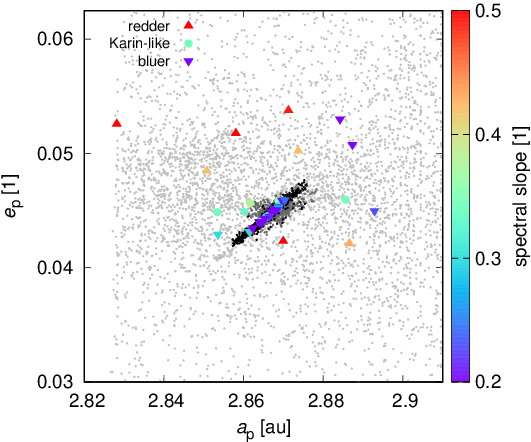}
\caption{
Positions of the Koronis family members in the proper element space,
which have observed spectral slopes
redder,
Karin-like,
or bluer,
compared to (832) Karin.
Red slope is an indication of a weathered surface and {\em vice versa\/}.
Bodies with less weathered Karin-like surfaces tend to be located close
to the young Karin and Koronis$_2$ families (black dots),
which is an indication that resurfacing is related to these families.
}
\label{vernazza_slope2_ae}
\end{figure}


\clearpage
\appendix

\section{Radiometric shock ages}\label{shock}

Measured shock ages of OCs also reveal non-uniform distributions
\citep{Swindle_2014GSLSP.378..333S}.
All mineralogical groups of OCs include numerous ages around 4560\,My,
most likely related to accretion.
A possible correspondence of other ages is summarized in Tab.~\ref{tab:shock}.

For H chondrites,
the oldest shock ages ${\gtrsim}\,3600\,{\rm My}$ might be related to Koronis,
which is the current source of H chondrites.
This is possible
if the original family is about 50\,\% older
and its SFD was about 50\,\% more populous.
The distribution of ages ${\lesssim}\,1000\,My$
seems to be continuous on a cumulative histogram.

For L-chondrite shock age 470\,My, there is a known candidate the Gefion family
\citep{Nesvorny_2009Icar..200..698N},
but its SFD is too shallow (cf. Tab.~\ref{ages}).
As discussed in \cite{Marsset_submit}, however,
the current source of shocked L chondrites is the the Massalia family,
suitably located in the inner main belt.

For LL chondrites,
the oldest ages ${\gtrsim}\,4200\,{\rm My}$ might be related to Eunomia.
Its SFD is shallow, but this sizable family is still a non-negligible source
of LL-chondrite-like material.
The sample of \cite{Swindle_2014GSLSP.378..333S} is too limited
to resolve details among younger shock ages.

Eventually, one cannot exclude the possibility that shocks originated also in
secondary collisional cascade,
minor cratering events,
microimpacts,
etc.

\begin{table}
\caption{
Possible correspondence of shock ages of OCs
and family-formation events.
}
\label{tab:shock}
\centering
\begin{tabular}{rll}
shock & family & \\
\vrule width 0pt depth 4pt
My & -- & \\
\hline
\vrule width 0pt height 9pt
3900  & Koronis (H) & if older \\
\\
 470  & Massalia (L) & if younger \\
\\
4200  & Eunomia (LL) \\
\hline
\end{tabular}
\end{table}


\section{Parameters of the principal bodies}\label{app:pb}

Hereinafter, we discuss preferred values of parameters
for the largest member of the studied families.
(4)~Vesta has a volume-equivalent diameter 525\,km
and a volumetric density $3.456\,{\rm g}\,{\rm cm}^{-3}$ \citep{Russell_2012Sci...336..684R};
the parent body size is practically the same as Vesta.

(20)~Massalia is 132\,km in diameter \citep{Usui_2011PASJ...63.1117U}
and its density is $3.71\,{\rm g}\,{\rm cm}^{-3}$ \citep{Carry_2012P&SS...73...98C},
although with a 20\% uncertainty.

For
(170)~Maria,
(808)~Merxia,
(847)~Agnia,
(158)~Koronis,
(1272)~Gefion,
we used diameters
35\,km,
33\,km,
30\,km,
34\,km \citep{Usui_2011PASJ...63.1117U}, and only
6.9\,km \citep{Nugent_2015ApJ...814..117N},
even though Gefion is not the largest remnant,
it has the lowest designation.
Because the densities are unknown, we assumed $3\,{\rm g}\,{\rm cm}^{-3}$.
All of these breakups were catastrophic disruptions;
parent body size is substantially larger, i.e.,
125\,km,
50\,km,
52\,km,
161\,km, and
72\,km, respectively.
This is important for the velocity field.
We determined these values by scaling of synthetic SFDs of \cite{Durda_2007Icar..186..498D};
uncertainties are of the order of 10\,\%

(832)~Karin is 14.3\,km in diameter
\citep{Usui_2011PASJ...63.1117U},
and the family parent body size is up to 36\,km.

For (3) Juno, we used
254\,km,
$3.15\,{\rm g}\,{\rm cm}^{-3}$,
according to \cite{Vernazza_2021A&A...654A..56V}.
It is the 2nd largest S-type asteroid.

For (8)~Flora,
146\,km,
$2.43\,{\rm g}\,{\rm cm}^{-3}$
from the same reference.
If about half of the family members has been lost in the $\nu_6$ resonance
the parent body size might have been larger.

(15) Eunomia is 270\,km in diameter,
and its density
$2.96\,{\rm g}\,{\rm cm}^{-3}$
is close the mean density of S-types
\citep{Usui_2011PASJ...63.1117U,Vernazza_2021A&A...654A..56V}.
It is the 1st largest S-type asteroid.

Finally, (44) Nysa is E-type, (135) Hertha is M-type;
both are likely interlopers in the respective S-type family.
Even without these interlopers, the parent-body size is up to about 80\,km,
as determined by the \cite{Durda_2007Icar..186..498D} method.


\section{Alternative family ages}\label{ages}

Previous orbital modelling, cratering record, or meteorite radiometry
can be used to estimate the age of an asteroid family.
The Vesta family is constrained by the Rheasylvia basin on Vesta,
or {\em in situ\/} observations
\citep{Marchi_2012Sci...336..690M,OBrien_2014P&SS..103..131O,Schenk_2022M&PS...57...22S},
as $(1000\pm 200)\,{\rm My}$.
It is in agreement with our collisional model (Tab.~\ref{tab3}).

The Phocaea family was studied by \cite{Carruba_2009MNRAS.398.1512C};
it is up to 2200\,My old, as inferred from the Yarkovsky drift rates.
Its SFD indicates a younger age (cf.~Tab.~\ref{tab3}).

The Massalia family is $(152\pm 18)\,{\rm My}$ old according to
\cite{Vokrouhlicky_2006Icar..182..118V}.
Parameters of the velocity field were also estimated,
$v_5 = 24\,{\rm m}\,{\rm s}^{-1}$,
$D_5 = 5\,{\rm km}$,
$v \propto D^{-1}$.
On contrary, its shallow SFD indicates an older age.

The Maria family may be up to $3000\pm1000\,{\rm My}$ old,
according to $(a_{\rm p}, H)$ distribution
\citep{Broz_2013A&A...551A.117B}.

The Merxia family, $(330\pm50)\,{\rm My}$ old
\citep{Vokrouhlicky_2006Icar..182..118V},
is almost certainly young,
having a smooth and steep SFD from the LR to the observational incompleteness.

The Agnia family is $(100\pm25)\,{\rm My}$ old
\citep{Vokrouhlicky_2006Icar..183..349V},
again smooth and steep.

On contrary, the Koronis family is really old,
$(2500\pm1000)\,{\rm My}$
\citep{Broz_2013A&A...551A.117B}.
Koronis is probably even older than Maria, because the `break' of the SFD
is at larger $D$'s (3 vs. 5\,km). 

The Gefion family was previously constrained by radiometry of LL chondrites
$(467\pm2)\,{\rm My}$
\citep{Greenwood_2007E&PSL.262..204G,Nesvorny_2009Icar..200..698N}.
On contrary, its SFD is shallow, which indicates an older age.

For the Juno family, we assume
$(750\pm150)\,{\rm My}$,
according to \cite{Carruba_2016MNRAS.457.1332C}.

The Flora family was estimated to be
$(1200\pm 200)\,{\rm My}$ old
\citep{Vokrouhlicky_2017AJ....153..172V}.
Our N-body modelling suggests that the synthetic family should be more extended,
with a substantially larger $D_{\rm PB} > 146\,{\rm km}$.
About half of bodies was lost in the $\nu_6$ resonance.

The Eunomia family is probably
$(3200\pm1000)\,{\rm My}$ old
\citep{Carruba_2016MNRAS.457.1332C}.
Our N-nody modelling suggests a range $1880$ up to $3300\,{\rm My}$
on the basis of the $(a_{\rm p},e_{\rm p})$ distribution.
It almost reaches a steady state,
because we recalibrate the synthetic SFD to the observed SFD in every time step,
which is then insensitive to the decay of the population
\citep{Broz_2013Icar..223..844B}.
Eunomia is most likely older than Flora (cf.~the `break').

Finally, the Nysa family is difficult to distinguish
from other overlapping families in the same region
\citep{Walsh_2013Icar..225..283W}.
S-type bodies are clustered around (135) Hertha
and the upper limit of its age is $350\,{\rm My}$
\citep{Dykhuis_2015Icar..252..199D}.


\section{Selection of `slow' shapes}\label{slow}

\begin{figure}
\centering
\includegraphics[width=9cm]{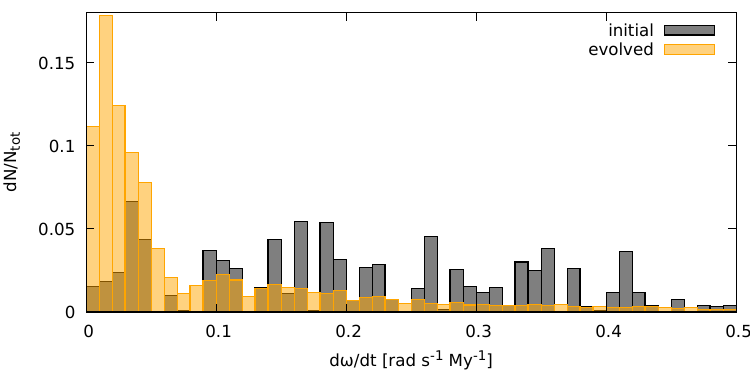}
\caption{
Normalized differential distribution $\d N/N_{\rm tot}$
of spin accelerations $\d\omega/\d t$
(in ${\rm rad}\,{\rm s}^{-1}\,{\rm My}^{-1}$)
for a population of metre-sizes bodies originated from the Agnia family;
close to the initial conditions (black)
and evolved due to the YORP effect ({\color{orange}orange\color{black}}).
A~strong preference for `slow' shapes is evident.
}
\label{agnia-4_1_slowshape_hist}
\end{figure}

In our orbital model, we noted a strong dynamical selection of shapes,
which evolve slowly due to the YORP torque
(Fig.~\ref{agnia-4_1_slowshape_hist}).
If the shape is 'fast',
the critical rotation frequency is reached fast,
this shape is changed to another one,
and {\em vice versa\/}.

\begin{figure*}
\centering
\includegraphics[width=12cm]{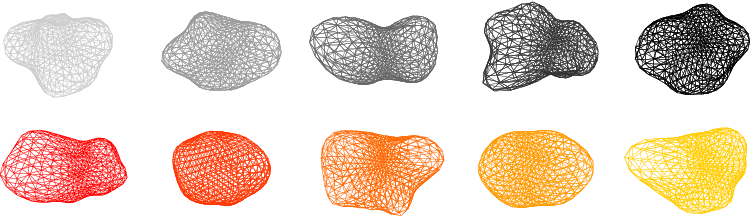}
\caption{
Examples of shapes from \cite{Capek_Vokrouhlicky_2004Icar..172..526C},
which exhibit fast (top) vs. slow (bottom)
evolution of the spin rate due to the YORP effect.
The latter appear to be more round, but it is generally difficult
to recognize a shape exhibiting a large vs. small YORP torque.
}
\label{agnia-4_1_gauss_meshes_slow}
\end{figure*}

Out of 200 nominal shapes from \cite{Capek_Vokrouhlicky_2004Icar..172..526C},
e.g., 185, 101, 129, 106, 58, \dots\ are slow (see Fig.~\ref{agnia-4_1_gauss_meshes_slow}).
They seem to be more round, but it is generally difficult to recognize it.
They should be less like a wind-mill \citep{Rubincam_2000Icar..148....2R}.

Moreover, the scaling relation we use in our model:
\begin{equation}
c = c_{\rm YORP}\left({a\over a_0}\right)^{-2}\left({R\over R_0}\right)^{-2}\left({\rho\over\rho_0}\right)^{-1},
\end{equation}
where
$a_0 = 2.5\,{\rm au}$,
$R_0 = 1\,{\rm km}$,
$\rho_0 = 2.5\,{\rm g}\,{\rm cm}^{-3}$,
is not complete.
A scaling with the rotation period (or frequency) is missing.
While the nominal period $P_0 = 6\,{\rm h}$,
for which the torques were originally computed,
is too long for meteoroids,
the YORP effect should work even in the limit of zero conductivity \citep{Rubincam_2000Icar..148....2R}.
It implies a negligible dependence on the rotation period.
This may change, if a transversal heat diffusion in mm- to cm-scale
surface features is properly taken into account
\citep{Golubov_2012ApJ...752L..11G,Sevecek_2015MNRAS.450.2104S}.
However, it would require a dedicated computation of the YORP effect
for metre-sized bodies.

\section{Supplementary tables}

The intrinsic collisional probability and the mean collisional velocity
were computed with \cite{Greenberg_1982AJ.....87..184G} theory
for precessing orbits.
The values for various combinations of populations are listed
in Tabs.~\ref{tabc1}, \ref{tabc2}.

\begin{table}
\caption{Intrinsic collisional probability and the mean collisional velocity for various main belt populations.}
\label{tabc1}
\centering
\begin{tabular}{lrr}
\vrule width 0pt depth 4pt
populations & $p$ & $v$ \\
--          & $10^{-18}\,{\rm km}^{-2}\,{\rm y}^{-1}$ & ${\rm km}\,{\rm s}^{-1}$ \\
\hline
\vrule width 0pt height 9pt
MB--MB                 &  2.860 &  5.772 \\
MB--Agnia              &  4.466 &  4.471 \\
MB--Eunomia            &  3.347 &  5.784 \\
MB--Flora              &  2.736 &  5.667 \\
MB--Gefion             &  3.545 &  5.115 \\
MB--Juno               &  3.009 &  6.491 \\
MB--Koronis            &  4.657 &  4.271 \\
MB--Maria              &  2.923 &  6.095 \\
MB--Massalia           &  4.269 &  5.042 \\
MB--Merxia             &  4.057 &  4.744 \\
MB--Nysa               &  3.986 &  5.093 \\
MB--Phocaea            &  2.419 &  8.252 \\
MB--Vesta              &  2.919 &  5.288 \\
Agnia--Agnia           & 10.535 &  2.241 \\
Eunomia--Eunomia       &  5.961 &  5.725 \\
Flora--Flora           & 15.506 &  4.235 \\
Gefion--Gefion         &  5.913 &  4.352 \\
Juno--Juno             &  4.950 &  7.034 \\
Karin--Karin           & 14.865 &  1.531 \\
Koronis--Koronis       & 13.323 &  1.625 \\
Maria--Maria           &  7.112 &  5.866 \\
Massalia--Massalia     & 29.009 &  4.234 \\
Merxia--Merxia         &  8.235 &  3.571 \\
Nysa--Nysa             & 20.324 &  4.766 \\
Phocaea--Phocaea       &  5.936 & 10.307 \\ 
Vesta--Vesta           & 12.601 &  3.613 \\
\hline
\end{tabular}
\end{table}

\begin{table}
\caption{Same as Tab.~\ref{tabc1} for the Earth and meteoroids in the NEO space.}
\label{tabc2}
\centering
\begin{tabular}{lrr}
\vrule width 0pt depth 4pt
populations & $p$ & $v$ \\
-- & $10^{-18}\,{\rm km}^{-2}\,{\rm y}^{-1}$ & ${\rm km}\,{\rm s}^{-1}$ \\
\hline
\vrule width 0pt height 9pt
Earth--Agnia      & 52.706 & 27.670 \\
Earth--Eunomia    & 15.807 & 29.691 \\
Earth--Flora      & 37.629 & 26.081 \\
Earth--Gefion     & 15.537 & 27.144 \\
Earth--Juno       & 18.813 & 28.350 \\
Earth--Karin      & 57.267 & 26.648 \\
Earth--Koronis    & 61.964 & 25.594 \\
Earth--Maria      & 15.476 & 32.093 \\
Earth--Massalia   & 72.563 & 26.014 \\
Earth--Merxia     & 16.391 & 24.346 \\
Earth--Nysa       & 79.017 & 25.823 \\
Earth--Phocaea    & 10.079 & 34.096 \\
Earth--Vesta      & 40.718 & 26.652 \\
\hline
\end{tabular}
\end{table}

\clearpage


\section{Supplementary figures}

We show the outcome of families identification procedure
and a preferred extent of the families in Fig.~\ref{aei2}.


\begin{figure*}
\centering
\begin{tabular}{c@{\kern0.1cm}c@{\kern0.1cm}c}
\kern0.5cm Phocaea (H) &
\kern0.5cm Maria (H) &
\kern0.5cm Merxia (H) \\
\includegraphics[width=6.0cm]{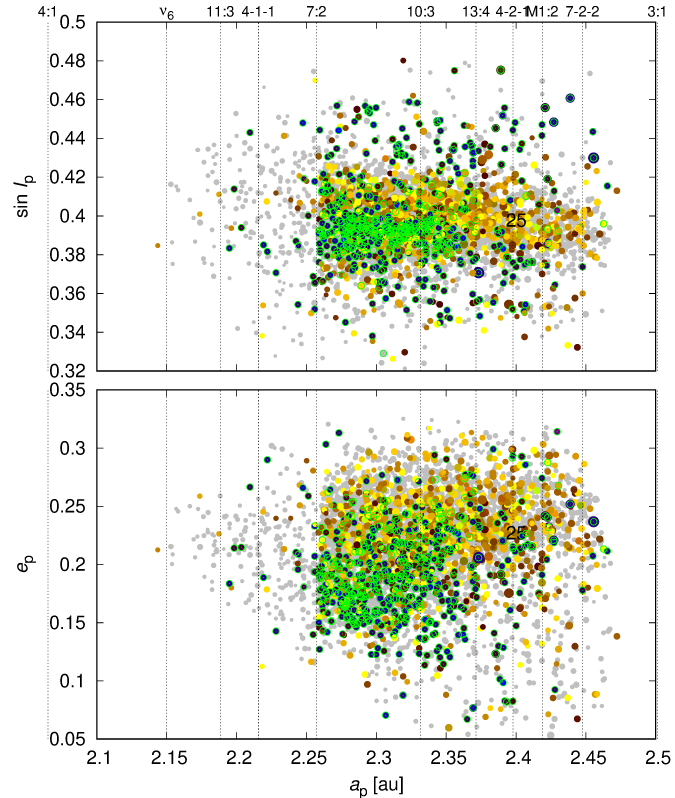} &
\includegraphics[width=6.0cm]{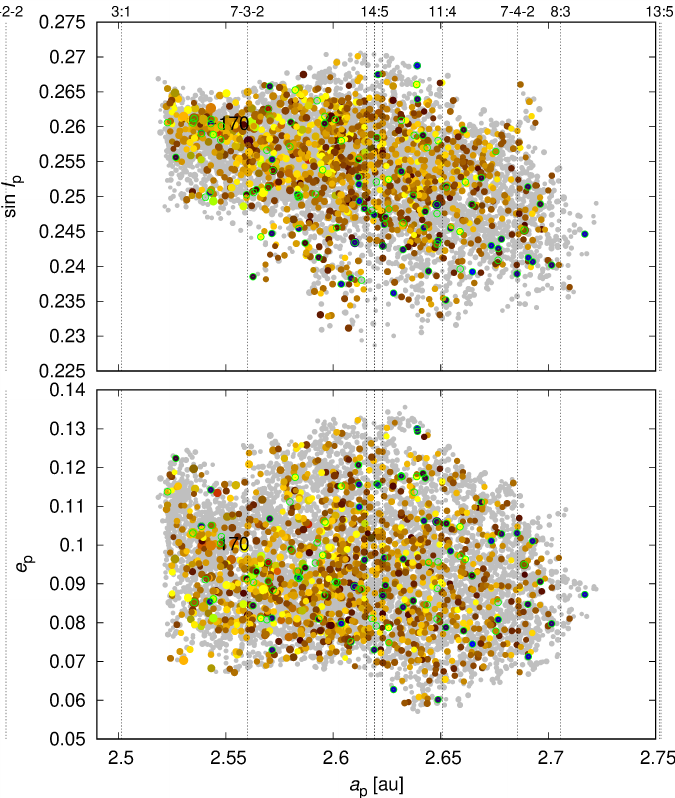} &
\includegraphics[width=6.0cm]{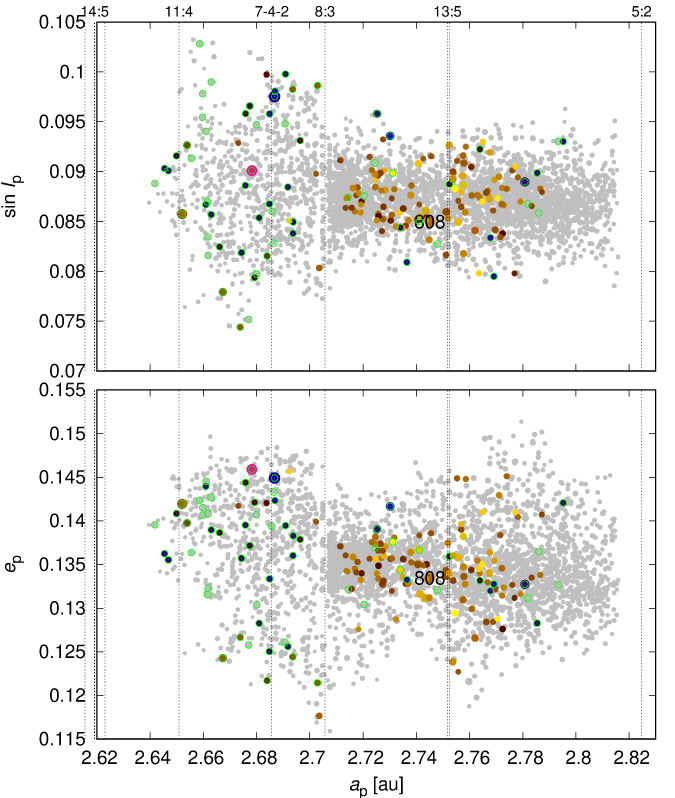} \\[0.1cm]
\kern0.5cm Agnia (H) &
\kern0.5cm Koronis (H) &
\kern0.5cm Massalia (L) \\
\includegraphics[width=6.0cm]{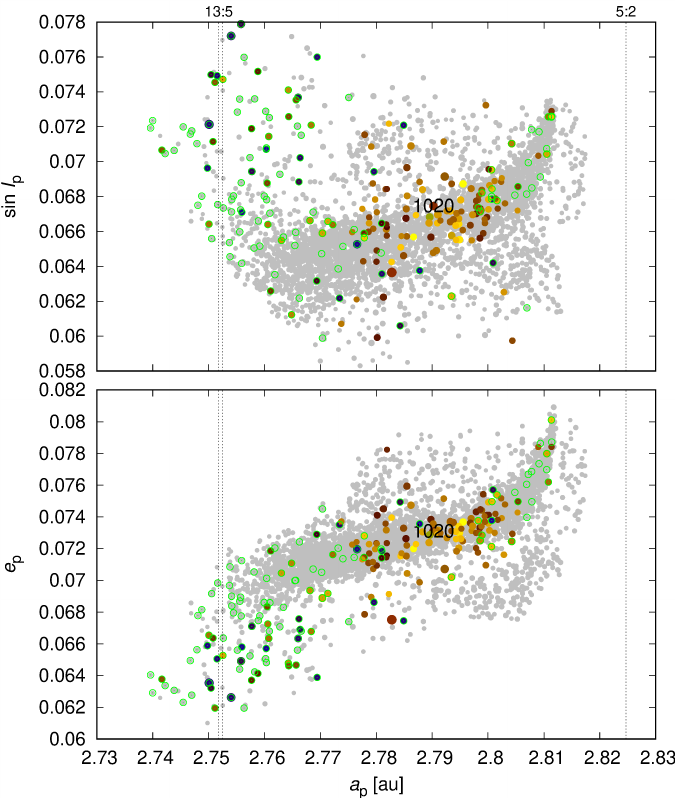} &
\includegraphics[width=6.0cm]{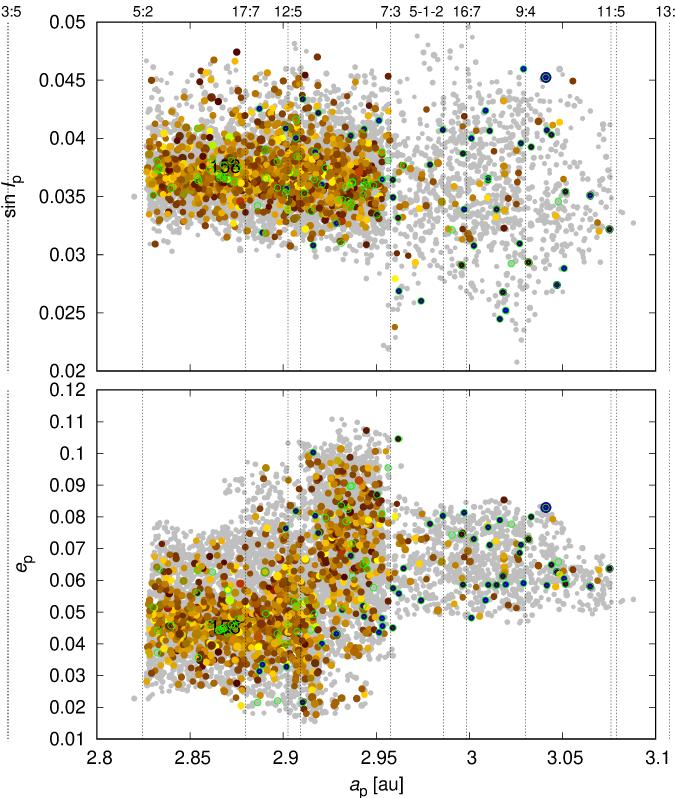} &
\includegraphics[width=6.0cm]{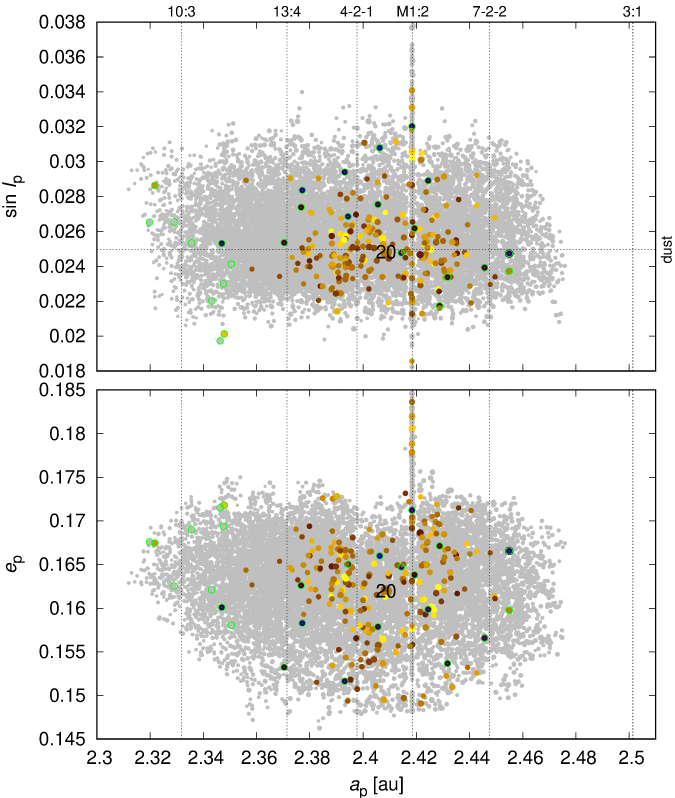} \\[0.1cm]
\end{tabular}
\caption{
S-type families as identified in this work.
The proper semimajor axis $a_{\rm p}$ vs. the proper eccentricity $e_{\rm p}$
and vs. the proper inclination $\sin I_{\rm p}$ are plotted.
Colours correspond to the geometric albedo~$p_{\rm V}$
(\color{blue}blue\color{black}$\,\rightarrow\,$\color{orange}yellow\color{black}\
for C- to S-type).
Major mean-motion and three-body resonances (vertical dotted lines),
as well as identified interlopers (\color{green}green\color{black}\ circles) are indicated.
Some of the bodies ((20) Massalia, (832) Karin) have inclinations
corresponding to the IRAS dust bands (horizontal dotted lines).
}
\label{aei2}
\end{figure*}

\addtocounter{figure}{-1}
\begin{figure*}
\centering
\begin{tabular}{c@{\kern0.1cm}c@{\kern0.1cm}c}
\kern0.5cm Gefion (L) &
\kern0.5cm Juno (L) &
\kern0.5cm Flora (LL) \\
\includegraphics[width=6.0cm]{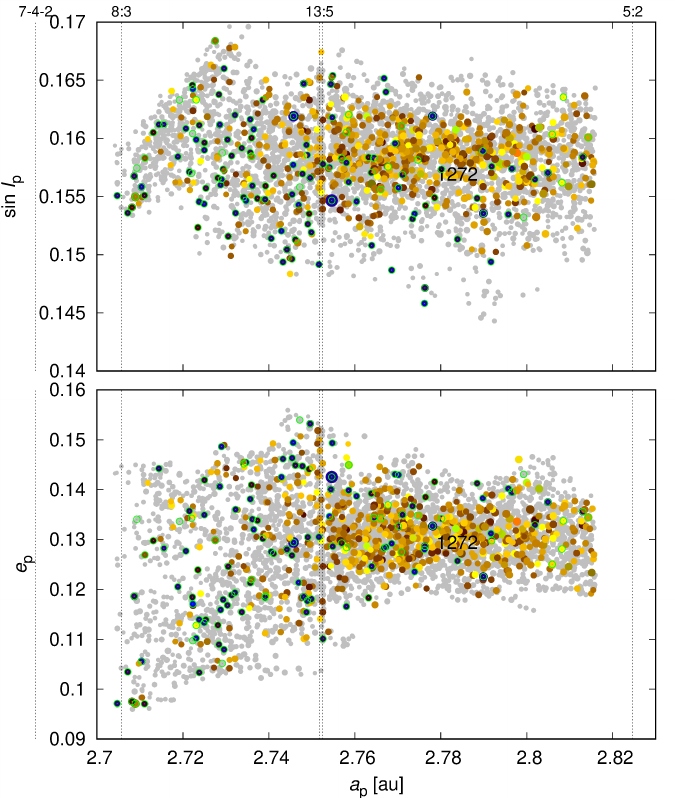} &
\includegraphics[width=6.0cm]{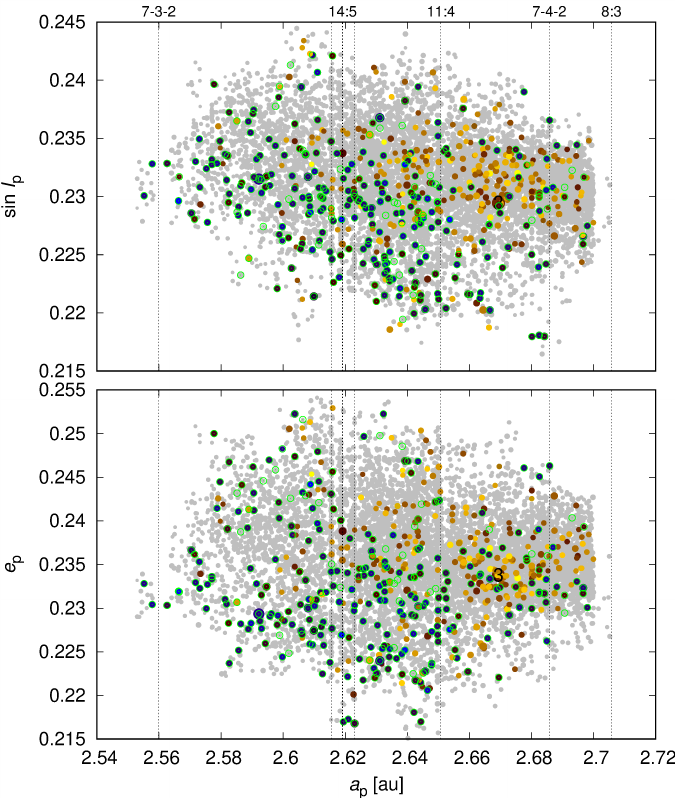} &
\includegraphics[width=6.0cm]{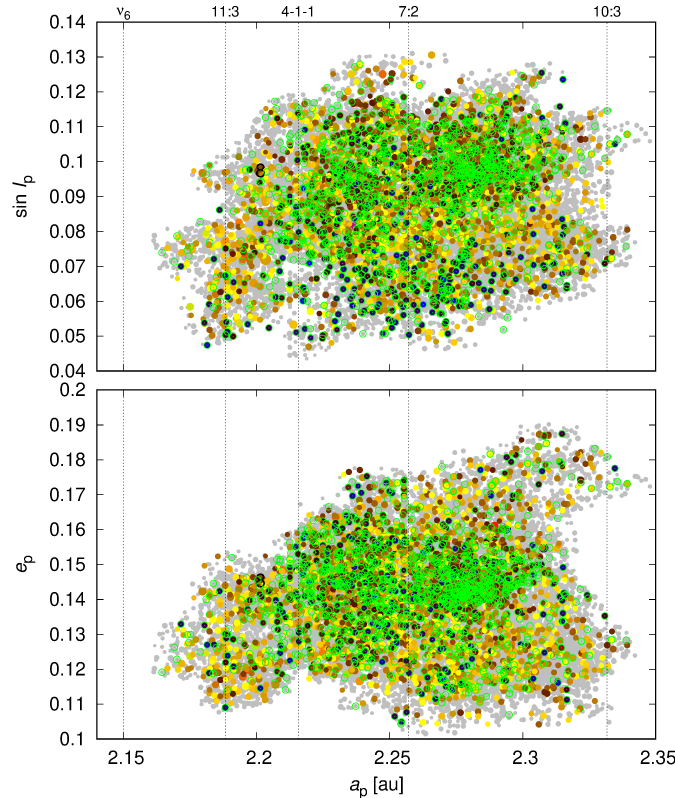} \\[0.1cm]
\kern0.5cm Eunomia (LL) &
\kern0.5cm Nysa (LL) &
\kern0.5cm Vesta (HED) \\
\includegraphics[width=6.0cm]{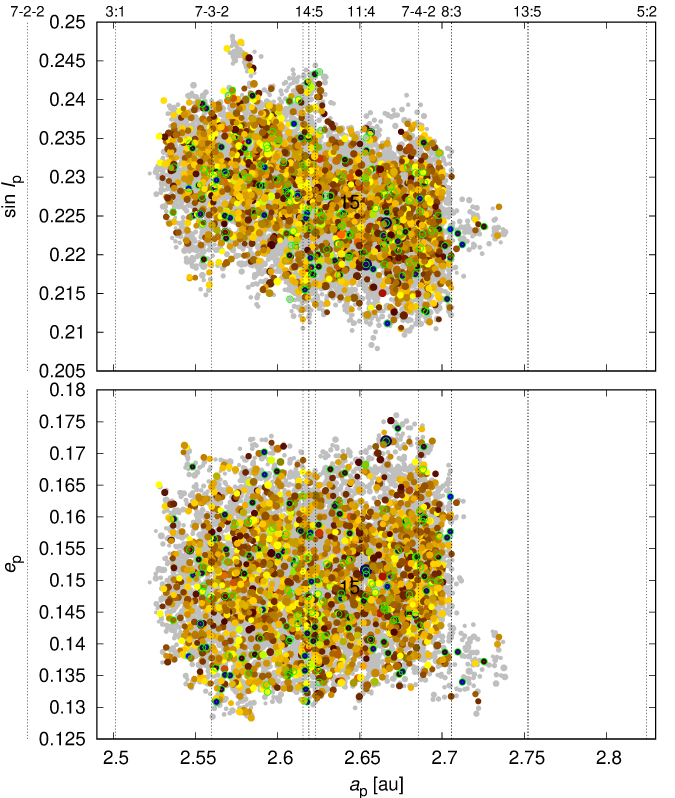} &
\includegraphics[width=6.0cm]{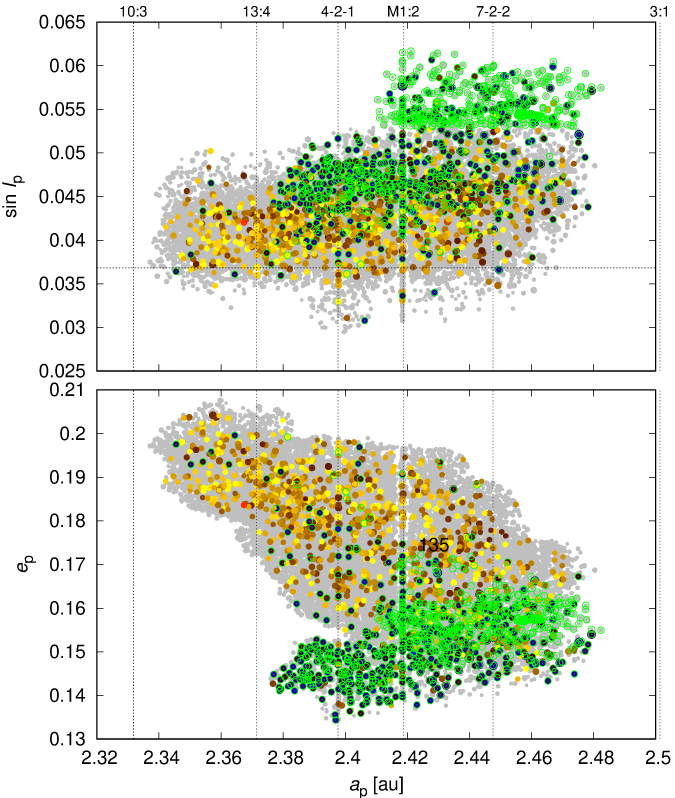} &
\includegraphics[width=6.0cm]{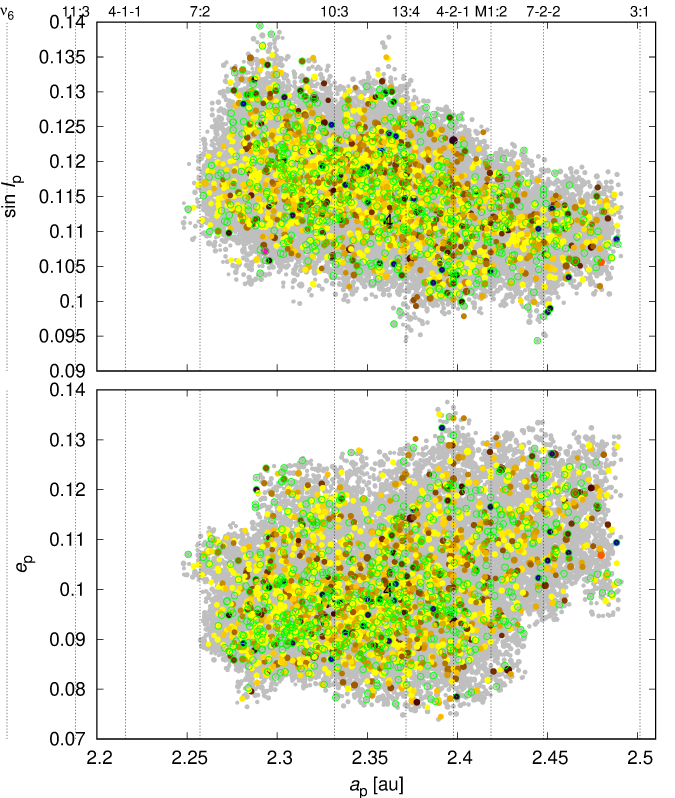} \\[0.1cm]
\end{tabular}
\caption{continued.}
\end{figure*}

\addtocounter{figure}{-1}
\begin{figure*}
\centering
\begin{tabular}{c@{\kern0.1cm}c@{\kern0.1cm}c}
\kern0.5cm Karin (H) &
\kern0.5cm Koronis$_{2}$ (H) \\
\includegraphics[width=6.0cm]{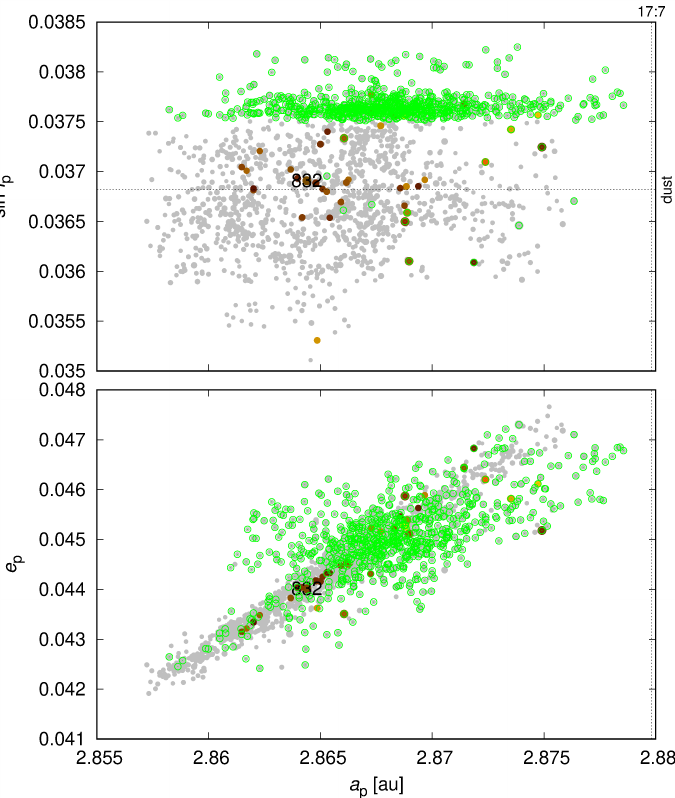} &
\includegraphics[width=6.0cm]{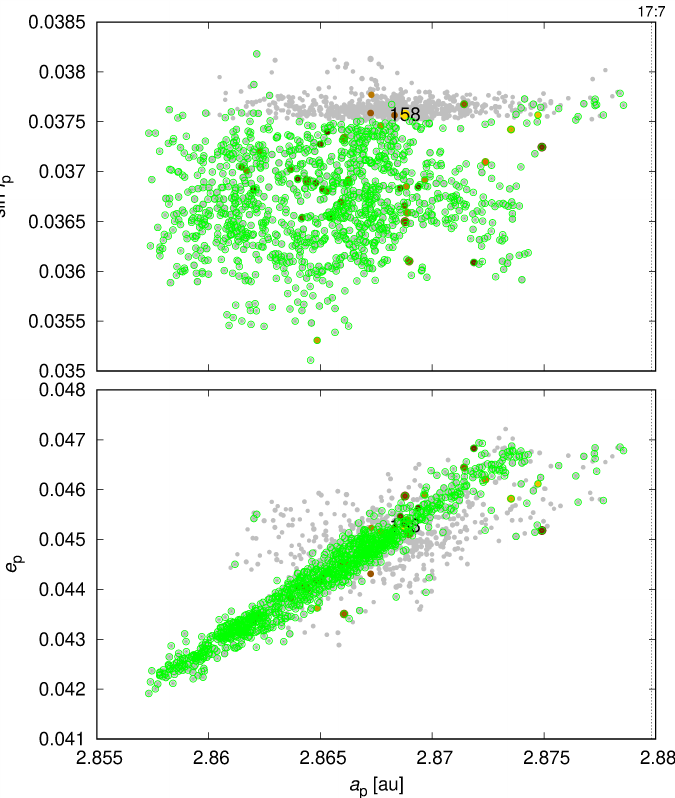} \\[0.1cm]
\end{tabular}
\caption{continued.}
\end{figure*}


\end{document}